\documentclass[conference]{IEEEtran}
\IEEEoverridecommandlockouts
\usepackage{cite}
\usepackage{amsmath,amssymb,amsfonts}
\usepackage{algorithmic}
\usepackage{graphicx}
\usepackage{textcomp}
\usepackage{xcolor}
\usepackage[hyphens]{url}

\definecolor{burgundy}{rgb}{0.5, 0.0, 0.13}
\definecolor{calpolypomonagreen}{rgb}{0.12, 0.3, 0.17}
\definecolor{deepsaffron}{rgb}{1.0, 0.6, 0.2}
\definecolor{vividauburn}{rgb}{0.58, 0.15, 0.14}
\definecolor{ultramarine}{rgb}{0.07, 0.04, 0.56}

\usepackage{fancyhdr}
\usepackage[bookmarks=true,breaklinks=true,letterpaper=true,colorlinks,citecolor=vividauburn,linkcolor=ultramarine,urlcolor=ultramarine]{hyperref}

\usepackage{tikz}
\usepackage[table]{xcolor}
\usepackage{pifont}
\usepackage{minted}
\usepackage{listings}
\usepackage{subcaption}
\usepackage{graphicx}
\usepackage{soul}
\usepackage{setspace}
\usepackage{xspace}
\usepackage{booktabs}
\usepackage[binary-units]{siunitx}
\usepackage{algorithm}
\usepackage{multirow}
\usepackage[most]{tcolorbox}

\def\BibTeX{{\rm B\kern-.05em{\sc i\kern-.025em b}\kern-.08em
    T\kern-.1667em\lower.7ex\hbox{E}\kern-.125emX}}
\begin{document}

% % Ensure letter paper
% \pdfpagewidth=8.5in
% \pdfpageheight=11in

%%%%%%%%%%%---SETME-----%%%%%%%%%%%%%
\newcommand{\iscasubmissionnumber}{135}
%%%%%%%%%%%%%%%%%%%%%%%%%%%%%%%%%%%%

% Helper to write numbers in a black circle.
\newcommand*\circled[1]{\tikz[baseline=(char.base)]{
            \node[shape=circle,fill,inner sep=0.9pt] (char) {\textcolor{white}{#1}};}}

\newcommand*\rect[1]{\tikz[baseline=(char.base)]{
            \node[shape=rectangle,fill,inner sep=1.5pt] (char) {\textcolor{white}{#1}};}}

\newcommand{\framework}{IP-CaT\xspace}
\newcommand{\frameworkfullname}{Instruction Prefetch Centric Cache and TLB Management (IP-CaT)\xspace}

\newcommand{\gv}[1]{\textcolor{orange}{(Georgios): #1}}
\newcommand{\mc}[1]{\textcolor{red}{(Marc): #1}}
\newcommand{\alex}[1]{\textcolor{olive}{(Alex): #1}}

\newcommand*\rev[1]{\textcolor{black}{#1}}
\newcommand*\revhpca[1]{\textcolor{black}{#1}}
\newcommand*\shp[1]{\textcolor{black}{#1}}

% Defining acronyms
\newcommand{\bcppr}{TIPRP\xspace}
\newcommand{\pcp}{PIP\xspace}
\newcommand{\npcp}{NPIP\xspace}
\newcommand{\bip}{BIP\xspace}

\newcommand{\tpb}{tPB\xspace}

\renewcommand{\sectionautorefname}{Section}
\renewcommand{\subsectionautorefname}{Section}
\renewcommand{\subsubsectionautorefname}{Section}

\pagenumbering{arabic}

%%%%%%%%%%%---SETME-----%%%%%%%%%%%%%
\title{Enhancing Instruction Prefetching\\via Cache and TLB Management}
% \title{\vspace{-0.24cm}Instruction Prefetch Centric\\Cache and TLB Management\vspace{-0.3cm}}

% \author{
%     \IEEEauthorblockN{\hspace{-1.5cm}Alexandre Valentin Jamet\IEEEauthorrefmark{1}%\vspace{0.065cm}
%     \IEEEauthorblockA{\hspace{-1.5cm}alexandre.jamet@bsc.es}\\
%     \IEEEauthorblockN{\hspace{3cm}Dimitrios Chasapis\IEEEauthorrefmark{1}}
%     \IEEEauthorblockA{\hspace{3cm}dimitrios.chasapis@bsc.es}}
%     \and
%     \IEEEauthorblockN{\hspace{-2.5cm}Georgios Vavouliotis\IEEEauthorrefmark{2}
%     \IEEEauthorblockA{\hspace{-2.5cm}georgios.vavouliotis@huawei.com}\\
%     \IEEEauthorblockN{\hspace{4.5cm}Marc Casas\IEEEauthorrefmark{1}}
%     \IEEEauthorblockA{\hspace{4.5cm}marc.casas@bsc.es}}
%     \and 
%     \IEEEauthorblockN{\hspace{-2.5cm}Mart\'i Torrents\IEEEauthorrefmark{4}\textsuperscript{\textdaggerdbl}
%     \IEEEauthorblockA{\hspace{-2.5cm}marti.torrents@bsc.es}\\
%     \IEEEauthorblockN{}
%     \IEEEauthorblockA{}}
%     \\
%     \IEEEauthorblockA{
%         \hspace{-14cm}\IEEEauthorrefmark{1}Barcelona Supercomputing Center%\hspace{1cm}
%         \IEEEauthorrefmark{2}Computing Systems Lab, Huawei Technologies AG, Switzerland
%         \IEEEauthorrefmark{4}Intel Corporation
%     }
% \vspace{-0.8cm}
% }

\author{
    \IEEEauthorblockN{\hspace{-1.5cm}Alexandre Valentin Jamet\IEEEauthorrefmark{1}%\vspace{0.065cm}
    \IEEEauthorblockA{\hspace{-1.5cm}alexandre.jamet@bsc.es}\\
    \IEEEauthorblockN{\hspace{3cm}Dimitrios Chasapis\IEEEauthorrefmark{1}}
    \IEEEauthorblockA{\hspace{3cm}dimitrios.chasapis@bsc.es}}
    \and
    \IEEEauthorblockN{\hspace{-3.5cm}Georgios Vavouliotis\IEEEauthorrefmark{2}
    \IEEEauthorblockA{\hspace{-3.5cm}georgios.vavouliotis@huawei.com}\\
    \IEEEauthorblockN{\hspace{2.5cm}Marc Casas\IEEEauthorrefmark{1}}
    \IEEEauthorblockA{\hspace{2.5cm}marc.casas@bsc.es}}
    \and 
    \IEEEauthorblockN{\hspace{-1.5cm}Mart\'i Torrents\IEEEauthorrefmark{1} %\textsuperscript{\textdaggerdbl}
    \IEEEauthorblockA{\hspace{-1.5cm}marti.torrents@bsc.es}\\
    \IEEEauthorblockN{}
    \IEEEauthorblockA{}}
    \\
    \IEEEauthorblockA{
        \hspace{-15cm}\IEEEauthorrefmark{1}Barcelona Supercomputing Center\hspace{0.3cm}
        \IEEEauthorrefmark{2}Computing Systems Lab, Huawei Technologies AG, Switzerland
        % \IEEEauthorrefmark{4}Intel Corporation
    }
\vspace{-0.8cm}
}

%%%%%%%%%%%%%%%%%%%%%%%%%%%%%%%%%%%%

\maketitle
\thispagestyle{plain}
\pagestyle{plain}

% \begingroup\renewcommand\thefootnote{\textdaggerdbl}
% \noindent \footnotetext{Work was done while the author was at Barcelona Supercomputing Center.}
% \endgroup

%%%%%%% -- PAPER CONTENT ENDS -- %%%%%%%%
% \begin{spacing}{0.99}
    \begin{abstract}
Modern server workloads have massive instruction footprints, exacerbating pressure on the processor front-end, and making techniques like L1 instruction (L1I) prefetching essential to alleviate this bottleneck. Although L1I prefetchers deliver significant performance gains, their full potential remains underutilized due to two main factors: i) L1I prefetch requests that cross page boundaries require address translation before being issued, and the latency involved in retrieving these translations undermines the timeliness of the L1I prefetches; ii) the reuse potential of code lines fetched in the cache hierarchy by L1I prefetches is highly variable—while a few lines are accessed multiple times, many are dead-on-arrival.

This paper proposes the \emph{\frameworkfullname}, a microarchitectural scheme orchestrating TLB and cache management to maximize the benefits of L1I prefetching. \framework comprises two modules: i) the \textit{translation Prefetch Buffer (tPB)}, a small buffer located alongside the last-level TLB (sTLB) that accommodates translations %page table entries (PTEs) 
fetched by L1I page-cross prefetches to  reduce the address translation cost of L1I prefetching and ii) the \textit{Trimodal Instruction Prefetch Replacement Policy (\bcppr)}, a decision-tree based replacement policy for the L2 cache (L2C) specialized in the management of lines fetched by L1I prefetches.

Our evaluation shows that \framework delivers significant performance benefits when integrated with three state-of-the-art L1I prefetchers (EPI~\cite{epi_isca}, FNL+MMA~\cite{fnlmma}, Barça~\cite{gratz2020barca}). For example, \framework{}+EPI achieves a \rev{6.1\%} geomean speedup over EPI across a set of 105 contemporary server workloads. We also show that \framework outperforms the state-of-the-art instruction TLB prefetcher~\cite{morrigan}, the leading TLB replacement policy~\cite{chirp}, and code-aware, prefetch-aware, and general-purpose cache replacement policies (Emissary~\cite{emissary}, SHiP++~\cite{Young2017SHiP}, Mockingjay~\cite{9773195}).
\end{abstract}


    \section{Introduction}
\label{sec:introduction}

\begin{figure}
    \centering
    \includegraphics[width=1\columnwidth]{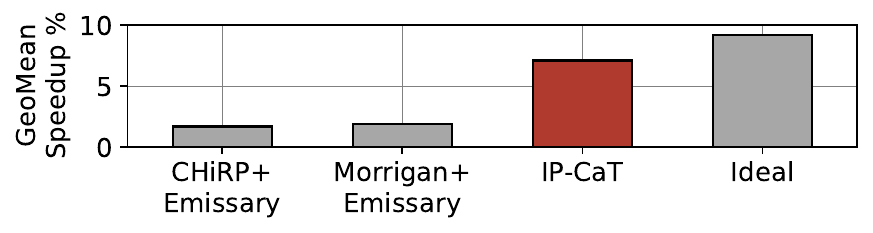}
    \vspace{-0.4cm}
    \caption{\rev{Comparison of \framework with i) combinations of the state-of-the-art TLB replacement policy (CHiRP~\cite{chirp}), instruction TLB prefetcher (Morrigan~\cite{morrigan}), and code-aware cache replacement policy (Emissary~\cite{emissary}) and ii) an idealized upper bound combining optimized TLB and cache management for L1I prefetches. This comparison considers EPI as L1I prefetcher~\cite{epi_isca} and 105 server workloads. }
    }
    \vspace{-0.2cm}
    \label{fig:introduction:epi_front_page_speedup}
\end{figure}

Contemporary server workloads feature massive instruction memory footprints. % that span multiple layers of complex software stacks. 
Prior studies~\cite{asmdb,bolt} reveal significant yearly growth in memory instruction footprints of these workloads, sometimes reaching up to 30\%. Although CPU vendors progressively increase the capacity of key front-end structures such as the L1 instruction cache (L1I) and the Translation Lookaside Buffer (TLB)~\cite{arm_neoverse_v2_chipsncheese}, these enhancements lag behind the rapid expansion of instruction working sets of server workloads~\cite{cloudsuite,bolt}. Consequently, the processor front-end faces increasing pressure, with frequent L1I misses stalling instruction fetching and degrading performance. In this context, microarchitectural techniques such as L1I prefetching have become indispensable to hide instruction fetch latencies and sustain high front-end performance.

L1I prefetchers~\cite{fdip,fnlmma,epi_isca,gratz2020barca} have proven effective in mitigating the front-end bottleneck of server workloads with large code footprints. These prefetchers use sophisticated mechanisms to identify instruction access patterns, enabling them to anticipate future instruction fetches and reduce costly cache misses. By proactively fetching instruction blocks before being demanded by the core, L1I prefetchers help sustain a steady instruction supply and improve overall pipeline utilization.

Modern L1I prefetching schemes operate with virtual addresses since first-level instruction caches are typically %implemented as 
Virtually Indexed and Physically Tagged (VIPT) structures~\cite{6237026,abishek_patterson_appendix}. Identifying and predicting instruction access patterns in the virtual address space is simpler than in the physical space, as adjacent virtual pages are not guaranteed to be contiguous in physical memory. %—an inconsistency that complicates prefetching. 
Operating in the virtual address domain also gives L1I prefetchers direct access to the TLB hierarchy, enabling them to issue prefetch requests that cross instruction page boundaries (hereafter, page-cross prefetch requests~\cite{a55,arm_neoverse_v2_chipsncheese,psa,vavou25}). An L1I prefetcher might be configured to discard or permit prefetches that cross page boundaries; the former is a conservative approach while the latter indirectly uses the L1I prefetcher to prefetch instruction translations to the TLB hierarchy~\cite{morrigan}. Allowing L1I prefetchers to cross page boundaries, a practice increasingly adopted in modern commercial processors~\cite{arm_neoverse_v2_chipsncheese}, enhances prefetchers' coverage and their ability to anticipate future instruction fetches.

Using three state-of-the-art L1I prefetchers~\cite{epi_isca,fnlmma,gratz2020barca}
%—EPI~\cite{epi_isca}, FNL+MMA~\cite{fnlmma}, and Barça~\cite{gratz2020barca}—
and a representative set of 105 server workloads, our analysis confirms that L1I prefetchers are highly effective, providing substantial performance gains. %In addition, our analysis identifies 
We identify two factors that undermine the potential of L1I prefetchers to deliver even higher performance. First, \emph{address translation latency becomes a critical bottleneck for L1I page-cross prefetches}. While L1I page-cross prefetching brings significant gains, % (\emph{e.g.}, higher coverage), 
its translation overhead offsets part of the benefit. We show that reducing the translation latency of  L1I page-cross prefetches reveals a significant opportunity for further performance gains. The second limiting factor is \emph{inefficient utilization of codes lines fetched by L1I prefetch requests in the lower-level caches}, particularly the L2 cache (L2C). Prefetch requests—both in-page and page-cross—fetch instruction lines with variable reuse behavior: many lines are dead-on-arrival, some experience limited reuse, and a small subset serve a large number of demand accesses. This variability indicates an opportunity to better coordinate L1I prefetching decisions with L2C management, thereby maximizing the utility of prefetched code lines.

We propose \emph{\frameworkfullname}, a microarchitectural scheme that jointly orchestrates TLB and cache management to amplify the benefits of L1I prefetching. \framework addresses the two key limitations that hinder the effectiveness of L1I prefetchers %, as revealed by our analysis, 
by i) mitigating the translation latency of L1I page-cross prefetching through the reuse of translations issued by the L1I prefetcher without polluting the TLB hierarchy and ii) reducing L2C pollution from useless L1I prefetches via a new decision tree-based replacement policy that anticipates the reuse potential of prefetched code lines. 

To amplify the benefits of L1I prefetching, \framework integrates two components. The first is the \textit{Translation Prefetch Buffer (tPB)}, a small buffer located alongside the TLB, which stores translations  fetched into the TLB hierarchy by L1I page-cross prefetches. The second is the \textit{Trimodal Instruction Prefetch Replacement Policy (\bcppr)}, a decision tree-based L2C replacement policy that combines three complementary policies to judiciously manage prefetched code lines by anticipating their reuse. \bcppr dynamically selects the most suitable policy per program phase based on a tree-based selection logic monitoring both cache hits and evictions. This policy goes beyond the monolithic, single-counter design of set-dueling~\cite{10.1145/1250662.1250709}, enabling more fine-grained and adaptive policy decisions. tPB accelerates L1I prefetching timeliness by hiding the address translation latency of page walks, while \bcppr minimizes L2C code pollution by preserving only the prefetched code lines with high reuse potential. Their synergistic interaction enables \framework to improve the performance of any L1I prefetcher by coordinating TLB and cache management for L1I prefetches. 

Our evaluation considers three state-of-the-art  L1I prefetchers to prove the versatility and effectiveness of \framework. Using a set of 105 contemporary server workloads, we show that \framework consistently enhances the performance of all evaluated L1I prefetchers by reducing the address translation latency of L1I page-cross prefetches and optimizing L2C management for code lines fetched by L1I prefetchers. As shown in Figure~\ref{fig:introduction:epi_front_page_speedup}, \framework significantly outperforms the state-of-the-art TLB replacement policy~\cite{chirp}, 
instruction TLB prefetcher~\cite{morrigan}, and code-aware cache replacement policy~\cite{emissary}, approaching the ideal upper bound of co-optimizing TLB and cache management for L1I prefetches. 

In summary, this paper makes the following contributions:

\begin{itemize}
    \item We show that instruction address translation limits the effectiveness of L1I page-cross prefetching and state-of-the-art schemes~\cite{morrigan,chirp} fail to exploit its full potential.

    \item We demonstrate that code lines fetched in L2C by L1I prefetchers exhibit highly variable reuse behavior, ranging from dead-on-arrival lines to heavily reused lines.

    \item We propose \emph{\framework}, the first microarchitectural scheme to fully extract the benefits of L1I prefetching through coordinated TLB and L2C management. % {\color{red} \framework integrates two components: the \textit{Translation Prefetch Buffer (tPB)} and the \textit{Trimodal Page-Cross Prefetch Replacement (\bcppr)}, which optimize TLB and L2C handling of translations and cache lines fetched by L1I prefetch requests, respectively.}
    
    \item We evaluate \framework using three state-of-the-art L1I prefetchers (EPI~\cite{epi_isca}, Barça~\cite{gratz2020barca}, FNL+MMA~\cite{fnlmma}) across 105 single-core server workloads and 160 multi-core server workload mixes. \framework consistently improves IPC across all considered L1I prefetchers and workloads. For example, when combined with EPI, \framework improves geomean performance by up to \rev{6.1\%}.

    \item We show that \framework outperforms the state-of-the-art instruction TLB prefetcher (Morrigan~\cite{morrigan}), TLB replacement policy (CHiRP~\cite{chirp}), and code-aware, prefetch-aware, and general-purpose cache replacement policies (Emissary~\cite{emissary}, SHiP++~\cite{Young2017SHiP}, Mockingjay~\cite{9773195}, CLIP~\cite{clip}, PACMAN~\cite{pacman}, \shp{PACIPV}~\cite{paciv}).
\end{itemize}

\section{Background on Cache Management}
\label{sec:background}
\label{subsec:background:cache_tlb_replacement}

Previously proposed cache replacement policies can be broadly classified in general-purpose policies, prefetch-aware policies, and code-aware policies.

\emph{General-purpose cache replacement policies} drive replacement decisions using i) the block recency without using histories of prior misses~\cite{Gao2010ADS,srrip,10.1145/301453.301487,10.1145/170035.170081,10.1145/1250662.1250709,4041862,824338} or (ii) features that correlate with the behavior of past accesses and anticipate the reuse distance of cache lines~\cite{ship,9773195,multiperspective,teran2016perceptron,5695535,glider,hawkeye,10.1109/MICRO.2012.43,10.5555/545215.545239,4601909,4358260,4771793,8091244}. The state-of-the-art general-purpose cache replacement policy for lower-level caches is \textit{Mockingjay}~\cite{9773195}, a scheme that uses patterns in long PC histories to accurately predict reuse distances.

\emph{Prefetch-aware cache replacement policies} distinguish between demand and prefetch requests to make replacement decisions. %SHiP++~\cite{Young2017SHiP} extends SHiP~\cite{ship} by deferring state updates until a prefetched line is reused. 
PACMan \cite{pacman} dynamically adjusts insertion and promotion policies to mitigate the negative impact of inaccurate prefetches. PACIPlV \cite{paciv} proposes a  replacement policy based on an offline exploration considering different insertion and promotion Re-Reference Prediction Values (RRPVs)~\cite{10.1145/2540708.2540733} for demand and prefetch lines. 

\emph{Code-aware cache replacement policies} target server applications with large code footprints and prioritize code lines over data lines in lower-level caches. The state-of-the-art policies in this category are CLIP~\cite{clip} and  Emissary \cite{emissary}. CLIP~\cite{clip} is built over the RRIP policy~\cite{srrip} and increases the priority of code lines in the L2C at the expense of having more data misses. Emissary \cite{emissary} prevents the eviction of the most critical for performance code blocks from L2C.

% The replacement policies for the TLB hierarchy are typically simple (\emph{e.g.}, LRU variants~\cite{colt,Saulsbury:2000,abishek_patterson_appendix}). The state-of-the-art TLB replacement policy is CHiRP \cite{9251943}, a predictive scheme that uses a signature-based prediction table to drive sTLB insertion decisions. To ensure correct update of its prediction table, CHiRP augments each sTLB entry with metadata. % bits. 
% \gv{I would defenitely remove this paragraph as well.}
% }
    \section{Motivation}
\label{sec:motivation}

This section focuses on the front-end bottleneck of server applications and motivates the need for new approaches that amplify the benefits of L1I prefetching. %Section~\ref{subsec:frontend_bottleneck} discusses the front-end bottleneck of server applications. 
\autoref{subsec:Instruction_prefetchers} reveals the potential benefits of reducing the translation latency of L1I page-cross prefetches. \autoref{subsec:motivation:impact_page_crossing_prefetch_l2c_replacement} shows that smart management of lines fetched in L2C by L1I prefetches can provide significant gains. Section \ref{subsec:conclusions_of_motivation_analysis} summarizes our findings. 

\subsection{Front-end Bottleneck}
\label{subsec:frontend_bottleneck}

Contemporary server workloads exhibit large instruction working sets spanning multiple software layers, posing significant challenges for the processor front-end~\cite{cloudsuite}. Prior studies~\cite{asmdb,morrigan,schall2024llbp} report that these workloads impose substantial front-end overheads and that their instruction footprints grow by up to 30\% annually. Although key front-end structures such as the L1I and TLB have increased in size~\cite{arm_neoverse_v2_chipsncheese}, this growth lags behind the escalating demands of modern servers. As instruction footprints increase, more instruction PTEs are required to map the working set, stressing both the first-level 
TLB (iTLB) and the last-level TLB (sTLB). The resulting increase in iTLB MPKI leads to more instruction translation requests to the sTLB, which typically caches both data and instruction PTEs~\cite{abishek_patterson_appendix}. This added pressure intensifies contention between data and instruction entries, raising sTLB miss rates and triggering more long-latency page walks. Consequently, front-end stalls—dominated by sTLB instruction misses and L1I cache misses—account for over 10\% of total execution cycles in industrial server workloads~\cite{Kanev:2015:PWC,bolt}. 

\subsection{Impact of L1I Prefetching on TLBs} 
\label{subsec:Instruction_prefetchers}

L1I prefetchers~\cite{fdip, fnlmma, epi_isca, gratz2020barca} typically operate with virtual addresses as L1I is typically VIPT and, therefore, these prefetchers can freely trigger prefetches that span instruction page boundaries~\cite{a55,arm_neoverse_v2_chipsncheese}. L1I prefetchers can be configured either to discard or allow page-crossing prefetches. The former approach is conservative, while the latter is more aggressive, enabling the prefetcher to anticipate instruction translations and prefetch them to the TLB~\cite{morrigan}. Allowing page-crossing prefetches, a growing industry trend~\cite{arm_neoverse_v2_chipsncheese}, enhances the prefetcher's ability to anticipate future instruction accesses. However, the need for address translations on page-cross prefetches limits the potential of L1I prefetching.

\begin{figure}
    \centering
    \includegraphics[width=1.0\columnwidth]{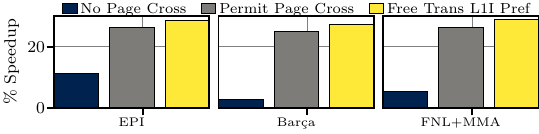}
    \vspace{-0.6cm}
    \caption{\rev{Geomean speedups when page-cross prefetches are discarded (No Page Cross), page-cross prefetching is permitted (Permit Page Cross), and an optimal scenario forcing sTLB hits for all L1I page-cross prefetches (Free Trans L1I Pref).}
    } 
    \vspace{-0.2cm}
    \label{fig:motivation:all_prefetcher_ideal_stlb}
\end{figure}

To quantify the advantages of allowing L1I prefetchers to cross page boundaries and the performance cost of translating these page-cross prefetches, we consider three state-of-the-art L1I prefetchers: EPI~\cite{epi_isca}, FNL+MMA~\cite{fnlmma}, and Barça~\cite{gratz2020barca}, deployed on top of a microarchitecture with a decoupled front-end and a 5-level radix tree page table~\cite{skylake_uarch}. \autoref{sec:methodology} presents the detailed experimental setup. \autoref{fig:motivation:all_prefetcher_ideal_stlb} evaluates three different scenarios for each L1I prefetcher: i) \textit{No Page Cross}, where the  prefetcher discards requests that cross page boundaries; ii) \textit{Permit Page Cross}, where the prefetcher issues all requests no matter if they cross page boundaries or not; and iii)  \textit{Free Translation L1I Prefetching}, where cross-page prefetches missing in sTLB are instantaneously converted in sTLB hits.  \autoref{fig:motivation:all_prefetcher_ideal_stlb} shows the speedups of the different L1I prefetchers and scenarios over the baseline (\autoref{table:champsim}) across a set of 105 server workloads, presented in \autoref{sec:methodology}.

Figure~\ref{fig:motivation:all_prefetcher_ideal_stlb} shows that enabling L1I prefetchers to cross page boundaries (\textit{Permit Page Cross}) consistently outperforms the \textit{No Page Cross} scenario. Moreover, the ideal scenario where L1I page-cross prefetches have no translation cost (\textit{Free Trans L1I Pref}) achieves additional performance gains over the \textit{Permit Page Cross} scenario. This behavior is observed across all evaluated L1I prefetchers. 
% For example, when considering the EPI prefetcher, \textit{Free Trans L1I Pref} achieves a 27.2\% speedup with respect to \textit{No Page Cross}, while \textit{PGC} improves \textit{No Page Cross} by 15\%, across all considered server workloads. We observe similar behavior for FNL+MMA and Barça.
The gap between \textit{Permit Page Cross} and \textit{Free Translation L1I Prefetching} arises because eliminating translation latency improves prefetching timeliness. Therefore, we conclude that, while allowing L1I prefetchers to cross pages significantly boosts performance for server workloads, the address translation overhead limits these benefits. In contrast to prior work showing that data prefetchers crossing pages can help or hurt depending on workload~\cite{vavou25}, our results indicate that L1I page-cross prefetching is \revhpca{mostly} beneficial. This is because instruction streams tend to follow sequential and loop-based control flow, making them more predictable than the often irregular patterns of data accesses.

% \vspace{0.3cm}
% \noindent\fcolorbox{black}{gray!10}{%
%     \parbox{\dimexpr\columnwidth-2\fboxsep-2\fboxrule\relax}{%
%         \emph{\textbf{\underline{Finding 1.} Allowing L1I prefetchers to cross page boundaries delivers performance gains. Reducing the instruction address translation costs for L1I page-cross prefetches has the potential to provide further performance enhancements.}}
%     }%
% }
% \vspace{0.1cm}

\begin{tcolorbox}[
    colback=gray!10,
    colframe=black,
    boxrule=0.5pt,
    arc=4pt,
    left=2pt,
    right=2pt,
    top=3pt,
    bottom=3pt
]
\emph{\textbf{Finding 1: Allowing L1I prefetchers to cross page boundaries improves performance. Reducing the instruction address translation cost of L1I page-cross prefetching  has the potential to provide additional performance gains.}}

\end{tcolorbox}

\subsection{Impact of L1I Prefetching on L2C Management} 
\label{subsec:motivation:impact_page_crossing_prefetch_l2c_replacement}

% This section evaluates the potential of optimizing the management of prefetched code lines in lower-level caches. We quantify the performance benefits of optimally managing lines inserted in L2C by L1I prefetches.\footnote{We don't focus on L1I nor LLC since we found larger headroom to optimize the management of prefetched code lines in L2C.} This study considers the EPI, FNL+MMA, and Barça prefetchers, similar to \autoref{subsec:Instruction_prefetchers}, configured to freely prefetch across page boundaries since \S \ref{subsec:Instruction_prefetchers} reveals that permitting L1I prefetchers to cross page boundaries improves IPC over the conservative scenario that discards them. 

This section evaluates the potential performance gains of optimizing the management of lines inserted in lower-level caches by L1I prefetches. This study targets the L2C and not the LLC since we found larger headroom to optimize the management of prefetched code lines in L2C. Regarding the L1I prefetchers, we consider the EPI, FNL+MMA, and Barça prefetchers, similar to \autoref{subsec:Instruction_prefetchers}, configured to freely prefetch across page boundaries since \autoref{subsec:Instruction_prefetchers} shows that permitting L1I prefetchers to cross page boundaries improves IPC over the conservative scenario that discards them.

\begin{figure}%[h]
    \includegraphics[width=1.0\columnwidth]{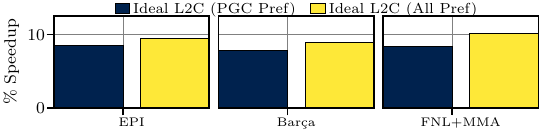}
    \vspace{-0.4cm}
    % \caption{\rev{Geomean speedups of an ideal scenario forcing L2C hits for lines fetched by L1I prefetches (Ideal L2C (All Pref)) and another ideal scenario forcing L2C hits for lines fetched by L1I page-cross prefetches (Ideal L2C (PGC Pref)).}
    \caption{\rev{Geomean speedups of two ideal scenarios forcing  L2C hits for lines fetched by i) L1I page-cross prefetches (Ideal L2C (PGC Pref)) and ii) L1I prefetches (Ideal L2C (All Pref)).} } 
    \vspace{-0.2cm}
    \label{fig:motivation:all_prefetcher_l2c_repl}
\end{figure}

\autoref{fig:motivation:all_prefetcher_l2c_repl} evaluates the impact of inserting code lines fetched by L1I prefetches in the L2C by considering two idealized scenarios: i) \rev{\textit{Ideal L2C (PGC Pref)}, where code lines fetched by page-cross prefetches are not inserted in L2C until a demand L2C access requests them. These entries are instead placed in an infinite buffer located alongside the L2C. When a demand access misses in L2C, we look up that buffer. If we have a hit, we magically insert that line in L2C.} This \rev{ideal} scenario quantifies how much performance can be extracted if code lines fetched by L1I page-cross prefetches incur no L2C pressure. ii) \textit{Ideal L2C (All Pref)}, where code lines fetched by L1I prefetches, both in-page and page-cross, are not inserted in L2C until a demand L2C access requests them. This scenario quantifies the performance if code lines fetched by L1I prefetches incur no L2C pressure.
The speedups in \autoref{fig:motivation:all_prefetcher_l2c_repl} are computed over the \textit{Permit Page Cross} version of each considered prefetcher in \autoref{fig:motivation:all_prefetcher_ideal_stlb}.

% \autoref{fig:motivation:all_prefetcher_l2c_repl} shows that the L2C pressure placed by L1I prefetches undermines their potential to improve performance. For example, EPI combined with \textit{Ideal L2C (All Pref)} scenario delivers a 8.0\% speedup over the \textit{Permit Page Cross} baseline. The main reason for the large performance gap between \textit{Permit Page Cross} and \textit{Ideal L2C (All Pref)} stems from the fact that a large number of prefetched code lines do not serve any demand L2C access. The \textit{Ideal L2C (All Pref)} speedups are significantly higher than the ones delivered by the \textit{Ideal L2C (PGC Pref)} scenario. The superior performance of \textit{Ideal L2C (All Pref)} with respect to \textit{Ideal L2C (PGC Pref)} indicates that judiciously managing prefetched code lines in L2C can bring larger performance improvements than only improving the management of L2C lines fetched by L1I prefetches that cross page boundaries.

\autoref{fig:motivation:all_prefetcher_l2c_repl} shows that the L2C pressure placed by L1I prefetches undermines their potential to improve performance. For example, EPI combined with \textit{Ideal L2C (All Pref)} scenario delivers a 9.5\% speedup over the \textit{Permit Page Cross} baseline. The main reason for the performance gap between \textit{Permit Page Cross} and \textit{Ideal L2C (All Pref)} stems from the fact that a large number of prefetched code lines do not serve any demand L2C access. The \textit{Ideal L2C (All Pref)} speedups are higher than the ones delivered by the \textit{Ideal L2C (PGC Pref)} scenario, highlighting that judiciously managing prefetched code lines in L2C can bring higher performance improvements than only improving the management of L2C lines fetched by L1I prefetches that cross page boundaries.

\subsubsection{Reuse of Prefetched Code Lines} 
\label{sss:reuseofprefetchedcodelines}

\begin{figure}
    \centering
    \includegraphics[width=1.0\columnwidth]{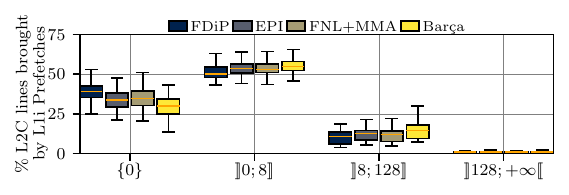}
    \vspace{-0.4cm}
    \caption{\rev{Breakdown of the number of demand L2C accesses served by code lines fetched in L2C by L1I prefetch requests. %\gv{there is room for increasing the height of the figure box given that the ylabel goes over it. I would try to optimize this figure for visibility.}
    }
    }
    \vspace{-0.6cm}
    \label{fig:motivation:dc_srv_l2c_hits}
\end{figure}

To support our claim that many prefetched code lines exhibit low reuse in L2C—and therefore harm performance—we measure how many demand L2C accesses are served by lines brought into L2C by L1I prefetches across the evaluated server workloads. \autoref{fig:motivation:dc_srv_l2c_hits} reports, for all L1I prefetchers as well as the baseline FDiP prefetcher~\cite{fdip}, the number of demand L2C accesses served by these prefetched lines. The x-axis groups prefetched L2C code lines by the number of demand accesses they serve, while the y-axis shows the fraction of lines in each group. %while the y-axis shows the portion of prefetched code lines in L2C belonging to each category.

\autoref{fig:motivation:dc_srv_l2c_hits} shows that all considered L1I prefetchers behave similarly when it comes to serving demand L2C accesses. Focusing on EPI, we observe that, on average, 36.1\% of prefetched code lines in the L2C remain unused, $i.e.$, serve no accesses, while 51.6\% serve between one and eight accesses. Additionally, 11.5\% of these lines handle more than eight accesses while 0.8\% of them serve more than 128 accesses during their time in L2C. The main takeaways of this study are that i) in many cases, L1I prefetch requests insert dead-on-arrival lines in L2C causing pollution (on average 36.1\%) and ii) a non-negligible fraction of the L1I prefetch requests bring code lines in L2C that serve a large number of demand L2C accesses, thus being very valuable for performance.

% \vspace{0.3cm}
% \noindent\fcolorbox{black}{gray!10}{%
%     \parbox{\dimexpr\columnwidth-2\fboxsep-2\fboxrule\relax}{%
%         \emph{\textbf{\underline{Finding 2.} Cache lines fetched in L2C by L1I prefetches show variable behavior, calling for a smart policy that anticipates the reuse of these prefetched code lines.}
%         }
%     }%
% }
% \vspace{0.2cm}

\begin{tcolorbox}[
    colback=gray!10,
    colframe=black,
    boxrule=0.5pt,
    arc=4pt,
    left=2pt,
    right=2pt,
    top=3pt,
    bottom=3pt
]
\emph{\textbf{Finding 2: Cache lines fetched in L2C by L1I prefetches show variable behavior, calling for a smart policy that anticipates the reuse of these prefetched code lines.}}
\end{tcolorbox}

We do not target the management of code lines fetched in L2C by demand accesses, as our analysis reveals low potential over prior art~\cite{emissary}. \autoref{subsec:evaluation:breakdown_epic_performance} shows that exclusively applying our proposal to prefetched code lines yields higher IPC than applying it to both demand and prefetch code lines.

% \begin{figure}%[h]
%     \centering
%     \includegraphics[width=1.0\columnwidth]{images/hpca_2026_final_ideal_comp_all_prefetcher_speedups.pdf}
%     \vspace{-1cm}
%     \caption{Geomean IPC speedups of previously studied scenarios in Figures~\ref{fig:motivation:all_prefetcher_ideal_stlb} and \ref{fig:motivation:all_prefetcher_l2c_repl} and an ideal scenario that combines them.} %, across  three L1I prefetchers (EPI, FNL+MMA, Barça).}
%     \vspace{-0.4cm}
%     \label{fig:motivation:final_ideal_comp_all_prefetcher_speedups}
% \end{figure}

\subsection{Putting Everything Together}
\label{subsec:conclusions_of_motivation_analysis}

Sections~\ref{subsec:Instruction_prefetchers} and \ref{subsec:motivation:impact_page_crossing_prefetch_l2c_replacement} highlight that contemporary L1I prefetchers deliver good performance gains. However, two factors undermine their potential for delivering outstanding benefits: i) the address translation latency of L1I page-cross prefetches and ii) the low reuse of a large fraction of code lines fetched in L2C by L1I prefetches. \autoref{sec:design} proposes a novel scheme that addresses our analysis findings and improves the performance of any L1I prefetching scheme.
    \section{\framework Design}
\label{sec:design}

This section presents \textit{\frameworkfullname}, the first scheme  to enhance the benefits of L1I prefetching through coordinated management of the TLB and the cache hierarchy. \framework comprises of two building blocks. First, the \textit{Translation Prefetch Buffer (tPB)}, a small buffer located alongside the sTLB that accommodates instruction page table entries (PTEs) fetched in the TLB hierarchy by L1I page-cross prefetches. tPB is motivated by Finding 1, which shows that while L1I page-cross prefetching can substantially improve the performance of server applications, its benefits are limited by instruction address translation latency. By enabling the reuse of PTEs fetched by the L1I prefetcher, tPB reduces the translation cost of L1I page-cross prefetching. Second, the \textit{Trimodal Instruction Prefetch Replacement Policy (\bcppr)}, a decision-tree L2C replacement policy specialized in the management of lines fetched by L1I  prefetches. \bcppr decides, at runtime, whether prefetched code lines in L2C should be retained, prioritized for eviction, or bypassed, based on their anticipated contribution to performance. The design of \bcppr is motivated by Finding 2, revealing that lines fetched in L2C by L1I prefetches shows variable behavior in terms of reuse; a big fraction of these lines are dead-on-arrival, some experience limited reuse, while a small fraction serves a large number of demand L2C accesses, thus being very critical for performance.

\subsection{\textit{translation Prefetch Buffer (tPB)}}
\label{subsec:translation_prefetch_buffer}

\begin{figure}%[h]
    \centering
    \includegraphics[width=0.605\columnwidth]{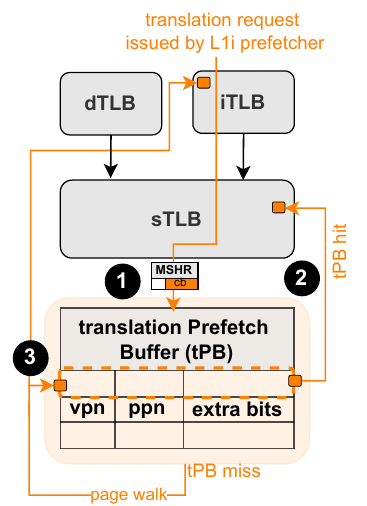}
    % \vspace{-0.1cm}
    \caption{Organization and operation of tPB.}
    \vspace{-0.2cm}
    \label{fig:translation_prefetch_buffer}
\end{figure}

The \textit{translation Prefetch Buffer (tPB)} is a small set-associative structure located alongside the sTLB that stores instruction PTEs fetched in the TLB hierarchy by L1I page-cross prefetches. Each tPB entry stores the virtual page number (\texttt{vpn}) for indexing purposes, the physical page number (\texttt{ppn}), and attribute bits that sTLB entries typically store~\cite{abishek_patterson_appendix}. Note that tPB is populated only by instruction translation requests originating from L1I page-cross prefetches and not by demand sTLB misses, justified in \autoref{subsec:evaluation:iso_storage_comp}. \autoref{fig:translation_prefetch_buffer} shows the design and operation (in steps) of tPB.

\emph{L1I prefetch requests} look up the TLB hierarchy (iTLB, sTLB) for the corresponding address translation. If the requested translation misses in both iTLB and sTLB, tPB is queried for possible hits (\circled{1} in \autoref{fig:translation_prefetch_buffer}). Upon tPB hits, the hitting tPB entry is inserted into sTLB \circled{2} following the sTLB insertion policy. Also, the hitting tPB entry is invalidated. For L1I prefetches that miss in tPB, a prefetch page table walk is initiated to fetch the corresponding address translation. Translation entries fetched by page walks initiated by the L1I prefetcher are stored in the iTLB and tPB, but not in the sTLB to avoid polluting the sTLB content \circled{3}.

\emph{Demand instruction TLB accesses} that miss in both iTLB and sTLB, are routed to the tPB. When a demand access hits in the tPB, the corresponding entry is inserted into the sTLB \circled{3} according to the sTLB insertion policy, under the assumption of future reuse. The matching tPB entry is then invalidated to free space in its limited storage. Therefore, tPB hits mitigate the address translation performance bottleneck by reducing the number of demand page table walks. \autoref{subsec:evaluation:perf_analysis} quantifies the impact of tPB on page walk reduction.

To support tPB's operation while allowing demand data and instruction translation requests to proceed normally, \framework should differentiate between translation requests originating from the L1I page-cross prefetches and the other translation requests. To do so, \framework requires one extra bit per sTLB MSHR entry indicating whether the corresponding translation request originates from an L1I page-cross prefetch request or not, as shown in \autoref{fig:translation_prefetch_buffer}. We refer to this bit as \emph{cross-bit (\texttt{cb})}. In practice, a new entry including the \texttt{cb} is inserted in the sTLB MSHR every time a translation request originating from an L1I page-cross prefetch misses in sTLB. The \texttt{cb} is set to 1 only for address translation requests coming from the L1I prefetcher and to 0 otherwise. The \texttt{cb} makes it possible to identify translations requested by L1I page-cross prefetches, which will be stored in tPB instead of sTLB (\circled{3} in \autoref{fig:translation_prefetch_buffer}) while all the other translations will be stored in the sTLB.

\begin{figure*}
    \centering
    \includegraphics[width=1\textwidth]{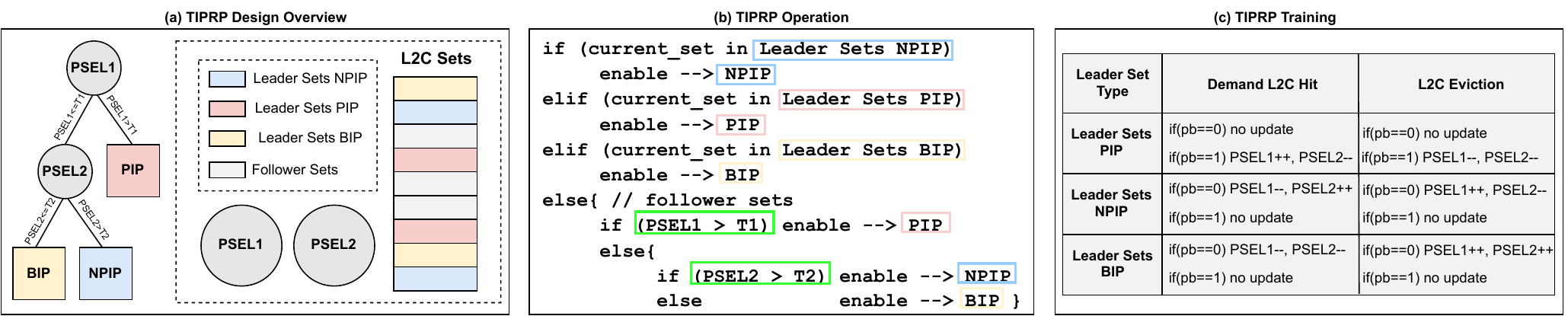}
    \vspace{-0.2cm}
    \caption{(a) Overview of \bcppr and the implementation of its adaptive selection logic that dynamically selects between \npcp and \pcp policies, (b) \bcppr operation in pseudo-code, and (c) \bcppr training events.} % \mc{Leader or Leader Sets?} \alex{Probably Leader.}}
    \vspace{-0.2cm}
    \label{fig:design:bcppr}
\end{figure*}

\subsubsection{\rev{Integrating tPB in sTLB}}
\label{sss:integration}

\rev{\autoref{subsec:translation_prefetch_buffer} presents tPB as a standalone structure to make its design and operation more transparent. 
In practical implementations, tPB can be seamlessly integrated into the sTLB by matching its associativity. %, \emph{i.e.}, using the same number of ways as the sTLB. 
Under this design, the sTLB is augmented with additional sets that are logically designated for tPB entries.} \rev{\autoref{subsec:evaluation:integrating_tpb_in_stlb} evaluates multiple tPB design alternatives and demonstrates that integrating tPB into the sTLB achieves performance gains comparable to those of a decoupled sTLB–tPB organization. Beyond simplifying the implementation, this integrated design also reduces the translation coherence overhead associated with a separate tPB structure, presented in \autoref{subsec:putting_it_together}.}

\subsection{\textit{Trimodal Instruction Prefetch Replacement Policy (\bcppr)}}
\label{subsec:l2c_bcppr}

The \textit{\underline{T}rimodal \underline{I}nstruction \underline{P}refetch \underline{R}eplacement (\bcppr)} is a decision tree-based L2C replacement policy that judiciously manages lines coming from L1I prefetches by anticipating whether these lines will be accessed in the future or not. \bcppr reduces L2C pollution incurred by dead-on-arrival lines fetched by L1I prefetches while maximizing the utilization of prefetched code lines that are critical for performance. 

\subsubsection{Building Blocks of \bcppr}
\label{subsubsec:design_overview_tiprp}

Since code lines fetched in L2C by L1I prefetches exhibit variable behavior (\autoref{subsec:motivation:impact_page_crossing_prefetch_l2c_replacement}), \bcppr combines three complementary RRPV-based~\cite{10.1145/1250662.1250709} policies. A decision tree dynamically selects between them, adapting \bcppr to the different execution phases (\autoref{subsubsec:dynamically_selecting_between_pcp_and_npcp}). \autoref{fig:design:bcppr} (a) illustrates the design of \bcppr\ along with its constituent replacement policies:

\textbullet~\emph{\underline{P}rioritize \underline{I}nstruction \underline{P}refetch (\pcp).} The \pcp policy protects lines fetched into L2C by L1I prefetches from being evicted. To select a candidate for eviction, \pcp looks  for the least recently used line not fetched to L2C by an L1I prefetch request present in the corresponding set.
If no such lines are present in the set, \pcp evicts the line in LRU position ($RRPV=3$). \pcp applies the standard SRRIP promotion and insertion policies ~\cite{10.1145/1250662.1250709, 5470352, 10.1145/2540708.2540733, chen_implementation_2006} for all cache lines.

\textbullet~\emph{\underline{N}on-\underline{P}rioritize \underline{I}nstruction \underline{P}refetch (\npcp).} \npcp favors the eviction of lines fetched by L1I prefetches. To do so, \npcp inserts lines fetched in L2C by L1I 
prefetches at the bottom of the recency stack, in the LRU position ($RRPV=3$). 
\npcp handles eviction and promotion of lines fetched by L1I prefetches in the same way as standard SRRIP~\cite{10.1145/1250662.1250709}. Finally, for all other cache lines types, \npcp applies the same eviction, promotion, and insertion policies as SRRIP~\cite{10.1145/1250662.1250709}.

\textbullet~\emph{\underline{B}ypass \underline{I}nstruction \underline{P}refetch (\bip).} \bip bypasses, $i.e.$, does not insert, code lines fetched in L2C by L1I prefetches. For other cache lines types, \bip applies the same eviction, promotion, and insertion policies as standard SRRIP~\cite{10.1145/1250662.1250709}.

% \noindent \emph{Insights on the operation of \pcp, \npcp, and \bip:} 
\subsubsection{Insights on the Operation of \pcp, \npcp, and \bip}
\label{subsubsec:fillme}

\pcp operates at eviction time while \npcp and \bip operate at insertion time. This asymmetry boosts \framework benefits and training efficiency. \pcp operates at eviction time to protect lines fetched by L1I prefetches. We found that applying it at insertion time would make it less effective in protecting prefetched code lines. In contrast, \npcp operates at insertion since this policy moderately favors the eviction of lines fetched into L2C by instruction prefetches. Operating at eviction time would bias \npcp too much towards the quick eviction of prefetched code lines. Finally, \bip completely avoids prefetched code lines to be inserted at L2C, thus can only operate at insertion time. 

\begin{figure*}
    \centering
    \includegraphics[width=1\textwidth]{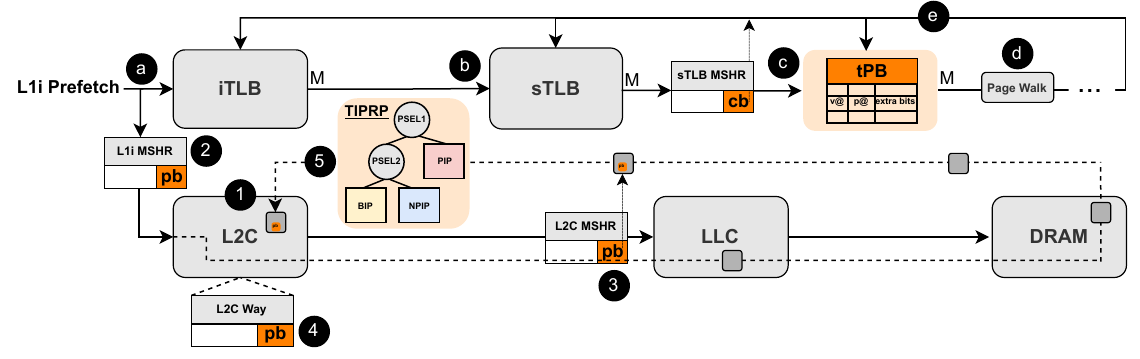}
    % \vspace{-0.1cm}
    \caption{\revhpca{Operation of \framework integrated in a standard microarchitecture.}}
    \vspace{-0.4cm}
    \label{fig:epic_design_schematic}
\end{figure*}

\subsubsection{Dynamically Selecting between \pcp, \npcp, and \bip}
\label{subsubsec:dynamically_selecting_between_pcp_and_npcp}
\bcppr dynamically decides whether \pcp, \npcp, or \bip should drive L2C eviction, promotion, and insertion policies. \bcppr goes beyond the monolithic nature of set-dueling~\cite{10.1145/1250662.1250709,casd_cal} %which is based on a single counter, 
by leveraging a two-level decision tree where node-level decisions are driven by saturating counters, as \autoref{fig:design:bcppr} (a) shows.
To determine the best policy for a given access, \bcppr uses two saturating counters, named \textit{PSEL1} and \textit{PSEL2}, that monitor the effectiveness of \pcp, \npcp, and \bip. %and will be used by the Follower Sets. 
\bcppr statically splits the L2C sets into four categories: i) Leader Sets for \pcp, \textit{i.e.}, sets statically assigned to use the \pcp policy, ii) Leader Sets for \npcp, \textit{i.e.}, sets statically assigned to use the \npcp policy, iii) Leader Sets for \bip, \textit{i.e.}, sets statically assigned to use the \bip policy, and iv) Follower Sets, which use the best policy between \pcp, \npcp, and \bip. 
Empirically, we determine that 10 bits for PSEL1 and PSEL2 and 32/16/16 Leader Sets for \pcp/\npcp/\bip are good decisions points. Note that \npcp and \bip use half as many Leader Sets as \pcp. Because \npcp and \bip do not prioritize lines fetched in L2C by L1I prefetches, the two can be seen as a single policy against \pcp. Using half Leader Sets for \npcp and \bip compared to \pcp ensures fair training and equal probability of selecting policies that either favor or do not favor lines 
fetched in L2C by L1I prefetches. %, giving the same weight to both classes of policies. 
\autoref{fig:design:bcppr} (a) presents an overview of \bcppr's selection logic. %\gv{maybe we want to add this: We use 32 Leader Sets for \pcp and half for each of the other two policies because \pcp favors prefetched code lines while \npcp and \bip assume that prefetched code lines pollute the cache, thus can be seen as a team.}

\framework requires one bit per L2C block to annotate whether a line has been fetched by an L1I prefetch or not. We refer to this bit as \textit{prefetch bit~(\texttt{pb})}. This bit makes it possible to update PSEL1 and PSEL2 counters, supporting the dynamic selection between the competing policies. \rev{Since \texttt{pb} is commonly present in L2C designs \cite{amd_prefetch_flag_2023,intel_sdm,arm_pl310_trm}, we assume the availability of \texttt{pb} at L2C, indicating whether a code line was fetched by the L1I prefetcher. For designs lacking this information, \autoref{subsec:putting_it_together} explains how \framework propagates \texttt{pb} to L2C.}

\textbullet~\textit{\bcppr Operation:} \autoref{fig:design:bcppr} (b) uses pseudo code to describe the operation of \bcppr. Upon an L2C access, \pcp, \npcp, or \bip drive the replacement for the current access based on whether or not the access belongs to either policy’s Leader Sets. If the access belongs to a Follower Set, PSEL1 and PSEL2 select which policy to activate. If PSEL1 is over  threshold $T{1}$, \pcp is selected. Otherwise, PSEL2 is used to make the final selection. If PSEL2 is below threshold $T{2}$, \bip is enabled; otherwise \npcp is used for the current access.

% \textbullet~\textit{Training of \bcppr's Selection Scheme:}
\textbullet~\textit{\bcppr Training:}
\autoref{fig:design:bcppr} (c) shows the training events of \bcppr's selection scheme, \emph{i.e.}, the update of the PSEL1 and PSEL2 counters. While original set-dueling \cite{10.1145/1250662.1250709} relies on a single counter and updates it only upon evictions, \bcppr updates PSEL1 and PSEL2 upon both L2C hits and L2C evictions while discriminating between cache lines fetched by L1I prefetches and the other lines. 

Regarding the \pcp training events, if a demand L2C request is served by a Leader Set of \pcp, PSEL1 and PSEL2 are updated only when the hitting line has been fetched in L2C by an L1I prefetch~(\texttt{pb}=1). In such case, PSEL1 is incremented (positive update), since hitting on L2C lines fetched by L1I prefetches in the \pcp Leader Sets indicates that \pcp is beneficial for performance. Conversely, evicting L2C lines fetched by L1I prefetches from \pcp Leader Sets implies that \pcp is not a good replacement policy for the current phase, thus PSEL1 is decremented (negative update), as shown in \autoref{fig:design:bcppr} (c). In both scenarios, PSEL2 is decremented, $i.e.$, we favor \bip over \npcp, since our experiments indicate that, when \bcppr determines that \pcp is not useful for the current program phase, it is better to aggressively start with \bip and gradually fall into \npcp if needed.

If a demand L2C request is served by a Leader Set of \npcp, PSEL1 and PSEL2 are updated only when the line that served the access has not been fetched into L2C by an L1I prefetch request (\texttt{pb}=0 in Figure  \ref{fig:design:bcppr} (c)). In this scenario, i) PSEL2 is incremented (positive update), as \autoref{fig:design:bcppr} (c) shows, since a hit on a L2C line not fetched by an L1I prefetch request in the \npcp Leader Sets indicates that \npcp brings benefits and ii) PSEL1 is decremented not to favor the selection of \pcp in subsequent accesses since \npcp brings benefits. Conversely, evicting L2C lines not fetched by an L1I prefetch request from \npcp Leader Sets indicates that \npcp is not optimally managing the L2C for the current phase, thus i) PSEL2 is decremented (negative update) to favor \bip over \npcp, $i.e.$, to promote the complete bypass of lines fetched by L1I prefetches and thus increase the L2C storage capacity dedicated to L2C lines not fetched by L1I prefetch requests; and ii) PSEL1 is incremented to favor the selection of \pcp for the next accesses. The operation for the Leader Sets of \bip is justified by similar arguments as \npcp.

\emph{Training Asymmetry.} \revhpca{Accesses to Leader Sets of \npcp and \bip update PSEL1 and PSEL2 only when the line has not been fetched in L2C by an L1I prefetch request (\texttt{pb}=0); when \texttt{pb}=1 no update happens. Similarly, accesses to Leader Sets of \pcp update PSEL1 and PSEL2 only when the line has been fetched in L2C by an L1I prefetch request (\texttt{pb}=1). The rationale of this asymmetry is that \bcppr updates PSEL1 and PSEL2 only on events that strongly indicate whether a policy improves or harms  performance. For example, upon an L2C eviction of a line not fetched by an L1I prefetch (\texttt{pb}=0) in an \npcp Leader Set, \framework updates PSEL1 to indicate that policies favoring the eviction of prefetched code lines are not producing the desired behavior, thus enabling \pcp as it can deliver better performance. Conversely, an L2C eviction of a line fetched by an L1I prefetch (\texttt{pb}=1) in an \npcp Leader Set does not carry much significance to decide whether \npcp should be enabled. We experimentally verify that updating PSEL1 and PSEL2 only on events strongly correlated to the usefulness of the considered policies, as shown in \autoref{fig:design:bcppr} (c), yields 5\% higher IPC than updating these counters in all events.} 

% \subsubsection{Insights on the Effectiveness of \bcppr}
% \label{}
\textbullet~\textit{Insights on the Effectiveness of \bcppr:} The placement of \pcp, \npcp, and \bip in the decision tree nodes of \autoref{fig:design:bcppr} (a) is crucial for the performance of \bcppr. We empirically determined that placing \pcp, \npcp, and \bip within the decision tree nodes in this way
allows for favoring transitions from \pcp to \npcp, and from \npcp to \bip. This configuration provides the highest performance improvement.

\subsection{Operation of \framework}
\label{subsec:putting_it_together}

\autoref{fig:epic_design_schematic} shows the complete operation of \framework. Standard microarchitectural structures appear in gray color while the components of \framework are annotated in different color. \rev{Since \texttt{pb} is typically present in L2C
designs~\cite{amd_prefetch_flag_2023,intel_sdm,arm_pl310_trm}, \framework does not require augmenting each L2C block with additional bits. \autoref{fig:epic_design_schematic} annotates \texttt{pb} in orange while showing how to propagate it to L2C for completeness.}

Upon L1I prefetch requests (for either in-page or page-cross prefetches), the iTLB is looked up for the corresponding address translation \circled{a} and, upon iTLB misses, the sTLB is accessed \circled{b}. For sTLB accesses that result in a miss, the tPB is looked-up for possible hits \circled{c}. Upon tPB hits, \framework inserts the requested translation in sTLB (\autoref{subsec:translation_prefetch_buffer}). Otherwise, a page walk is triggered to fetch the translation from the page table \circled{d}. At the end of the page walk \circled{e}, the requested translation is stored in i) tPB and iTLB for page walks triggered by L1I page-cross prefetch requests, and ii) iTLB and sTLB for page walks not triggered by  L1I page-cross prefetch requests, as explained in \autoref{subsec:translation_prefetch_buffer}. \framework is aware of whether translation requests originate from the L1I prefetcher or not since 
\revhpca{the sTLB MSHR} stores the \texttt{cb} bit, as described in \autoref{subsec:translation_prefetch_buffer} and \autoref{fig:epic_design_schematic}. \revhpca{Note both sTLB and tPB are looked up for potential hits upon requests for instruction translations, as explained in \autoref{subsec:translation_prefetch_buffer}.}

After address translation, the physical address is known and memory requests lookup the L2C upon first-level cache misses \circled{1}. \autoref{subsec:l2c_bcppr} explains how \bcppr drives the L2C insertion, promotion, and eviction policies. \framework decides to enable \npcp, \pcp, or \bip \circled{5} by taking into account the PSEL1 and PSEL2 values, as described in \autoref{subsec:l2c_bcppr}. \framework requires keeping \texttt{pb} in the L1I and L2C MSHRs (\circled{2},\circled{3}) to propagate \texttt{pb} to the L2C \circled{4} for designs lacking this feature. 

\textbullet~\emph{Storage Overhead.} \framework's storage overhead depends on the  sTLB MSHR size of the underline microarchitecture. \revhpca{Considering the system described in \autoref{sec:methodology}, \framework requires 0.79KB to be implemented (6452b for a 64-entry tPB, 16b for sTLB MSHR \texttt{cb} bits, \rev{10b for \texttt{PSEL1}, and 10b for \texttt{PSEL2}).}} This is just 0.08\% of the L2C capacity. The energy impact of \framework is negligible due to this minimal storage overhead.

\textbullet~\revhpca{\emph{TLB Shootdowns \rev{and Translation Coherence}.}} \revhpca{\framework does not introduce any translation coherence issue \rev{as the tPB component can be resembled as an extra TLB level or can be incorporated into the sTLB. The only requirement is to include tPB in the TLB shootdown process.} \rev{\autoref{sss:integration} describes that tPB can be seamlessly integrated in the sTLB, minimizing its translation coherence overheads. \autoref{subsec:evaluation:integrating_tpb_in_stlb} evaluates different tPB configurations suitable for sTLB integration.}}

\textbullet~\shp{\emph{Wrong-Path Execution.} \framework handles wrong-path requests implicitly, like conventional replacement policies that are agnostic to path correctness. While such requests may cause pollution or bring useful data, making replacement policies wrong-path aware is a promising direction for future work, e.g., via branch prediction hints or a dedicated predictor using commit-stage information.}

    \section{Experimental Methodology}
\label{sec:methodology}

\begin{table}%[ht]
    \centering
    \setlength\tabcolsep{2pt} % Adjust column padding
    \renewcommand{\arraystretch}{0.9} % Adjust row height
    \resizebox{1.0\columnwidth}{!}{ % Resize to fit column width
        \begin{tabular}{|l|m{9.2cm}|}
            \hline
            \rowcolor{gray!30} \textbf{Component} & \textbf{Description} \\
            \hline\hline
            \textbf{CPU Core} & 1- and 4-core system, 4GHz, 128-entry FTQ, 352-entry ROB, 6-wide issue \\
            \hline
            \rev{\textbf{Branch Predictor}} & \rev{TAGE-SC-L~\cite{SeznecM06,tage}, 8K-entry BTB, 64-entry RAS} \\ %hashed-perceptron~\cite{10.1145/1089008.1089011} \\
            \hline
            \textbf{L1I TLB (iTLB)} & 64-entry, 4-way, 1cc, \shp{8-entry MSHR}, LRU \\
            \hline
            \textbf{L1D TLB (dTLB)} & 64-entry, 4-way, 1cc, 8-entry MSHR, LRU \\
            \hline
            \textbf{L2 TLB (sTLB)} & 1536-entry, 12-way, 8cc, 16-entry MSHR, LRU \\
            \hline
            % \textbf{\textit{tPB}} & \textbf{0.79KB, 64-entry, fully-associative, LRU} \\
            % \hline
            \textbf{Page Structure} & 4-level Split PSC, parallel search, 1cc. \\
            \textbf{Caches (PSCs)} & L5: 1-entry, L4: 2-entry, L3: 8-entry, L2: 32-entry \\
            \hline
            \textbf{L1I Cache} & 32KB, 8-way, 4cc, \shp{8-entry MSHR}, LRU, EPI~\cite{epi_isca} / Barça~\cite{gratz2020barca} / FNL+MMA~\cite{fnlmma} \\
            \hline
            % \textbf{\textit{iPF}} & \textbf{4KB, 8192-entry prediction table, 4-bits counters} \\
            % \hline
            \textbf{L1D Cache} & 48KB, 12-way, 5cc, 16-entry MSHR, LRU, Berti \cite{berti} \\
            \hline
            \textbf{L2 Cache} & 1MB, 16-way, 10cc, 32-entry MSHR, LRU \\
            \hline
            \textbf{LLC} & 1.375MB per core, 11-way, 36cc, 64-entry MSHR, SHiP~\cite{ship} \\ %/56
            \hline
            % \textbf{DRAM} & \SI{4}{\giga\byte}/core, \SI{25.6}{GB/s} per core, 1 channel/core, $t_{RP}=t_{RCD}=t_{CAS}=\;$\SI{12.5}{ns} \\
            % \textbf{DRAM} & \SI{4}{\giga\byte}/core, \SI{25.6}{GB/s}, 1 channel/core, t\textsubscript{RP}=t\textsubscript{RCD}=t\textsubscript{CAS}=\;$\SI{12.5}{ns} \\
            \textbf{DRAM} & 4GB/core, 25.6GB/s, 1 channel/core, t\textsubscript{RP}=t\textsubscript{RCD}=t\textsubscript{CAS}=12.5ns \\
            \hline
        \end{tabular}
    }
    % \vspace{-0.1cm}
    \caption{Baseline System Configuration. %\gv{add here l1i prefetchers, remove no prefetcher}
    %\gv{we need to state which prefetchers we use for L1D, L2C.} \alex{it's there already.}
    }
    \vspace{-0.4cm}
    \label{table:champsim}
\end{table}

\begin{figure*}
    \centering
    \includegraphics[width=1\textwidth]{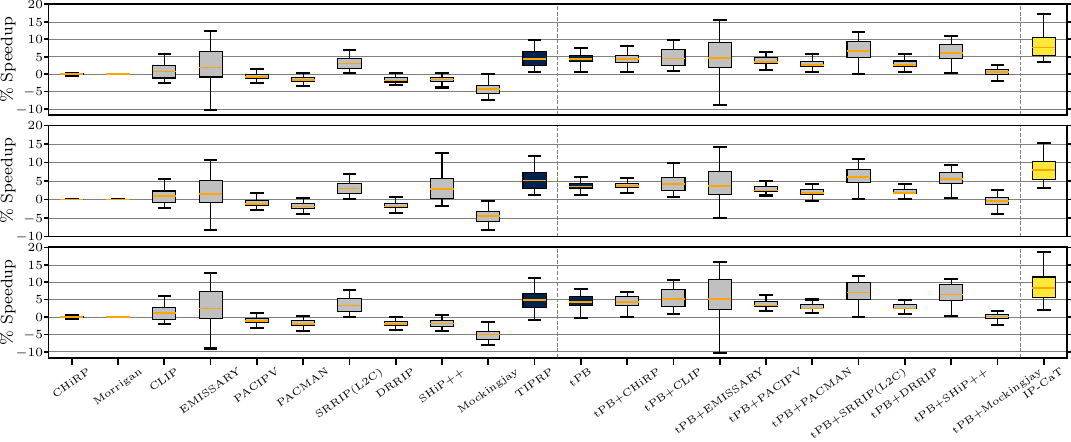}
    \vspace{-0.6cm}
    \caption{\rev{Evaluation considering either EPI (top), Barça (middle), or FNL+MMA (bottom) as L1I prefetcher.}
    }
    \vspace{-0.4cm}
    \label{fig:evaluation:all_speedups}
\end{figure*}

We evaluate \framework using ChampSim~\cite{champsim_cite,champsim_paper}, a detailed trace-based simulator of an out-of-order processor with a three-level cache hierarchy~\cite{berti}, and a decoupled front-end~\cite{decoupled_frontend_arm} with FDIP~\cite{fdip}. We consider 5-level radix tree page table, an x86 hardware page table walker~\cite{skylake_uarch}, and MMU Caches~\cite{10.5555/77493}. Table~\ref{table:champsim} details our baseline system configuration, \rev{similar to an Intel Cascade Lake microarchitecture~\cite{noauthor_cascade_nodate, 10476485}.}

% {\it L1I Prefetchers.} 
% We consider three state-of-the-art L1I prefetchers: FNL+MMA \cite{fnlmma}, Barça \cite{gratz2020barca}, and EPI \cite{epi_isca}. \shp{All L1I prefetchers are configured to prefetch freely across page boundaries, as \S \ref{subsec:Instruction_prefetchers} shows that allowing L1I page-cross prefetching provides significant gains compared to the approach that disables it.} 
% \gv{I would remove/merge this given 1st paragraph of the evaluation.}{\it L1I Prefetchers.} We consider three state-of-the-art L1I prefetchers: FNL+MMA \cite{fnlmma}, Barça \cite{gratz2020barca}, and EPI \cite{epi_isca}. \shp{All are configured to prefetch across page boundaries, as \S \ref{subsec:Instruction_prefetchers} shows this yields significant performance gains over disabling it.}

{\it Simulated Page Sizes.} %\framework can be seamlessly applied in systems with different page sizes. 
Our evaluation considers two scenarios: i) the system uses only 4KB pages (\autoref{subsec:overall},  \autoref{subsec:evaluation:breakdown_epic_performance}) and ii) the system uses both 4KB pages and 2MB pages (\autoref{subsec:evaluation:multi_page_size_performance_analysis})~\cite{psa,10.1145/3669940.3707247}. %We use the methodology proposed by prior work~\cite{psa,victima,10.1145/3669940.3707247} in the multi-page size scenario. 
Considering 4KB pages is relevant as large pages require memory contiguity and defragmentation, % to deliver competitive performance, 
which cannot be guaranteed in servers due to their large uptimes~\cite{10.1145/3669940.3707247,asmdb,osdi21,ranger}. 

\begin{table}%[ht]
    \centering
    \setlength\tabcolsep{4pt} % Adjust column padding
    \renewcommand{\arraystretch}{0.9} % Adjust row height
    \resizebox{1.0\columnwidth}{!}{ % Resize to fit column width
        \begin{tabular}{|l|m{3.0cm}|m{3.0cm}|c|c|}
            \hline
            \rowcolor{gray!30} \textbf{Technique} & \centering \textbf{L2C} & \centering \textbf{LLC} & \textbf{STLB} & \textbf{tPB} \\
            \hline
            \hline

            \rowcolor{gray!45} \rev{Baseline} & \centering LRU & \centering \rev{SHiP} & \rev{LRU} & \rev{n/a} \\
            
            \hline\hline
            CHiRP~\cite{chirp} & \centering LRU & \centering \rev{SHiP} & CHiRP & n/a \\
            \hline
            Morrigan~\cite{morrigan} & \centering LRU & \centering \rev{SHiP} & Morrigan & n/a \\
            \hline
            CLIP~\cite{clip} & \centering CLIP & \centering SHiP & LRU & n/a \\
            \hline
            EMISSARY~\cite{emissary} & \centering EMISSARY & \centering SHiP & LRU & n/a \\
            \hline
            PACIPV~\cite{paciv} & \centering LRU & \centering PACIPV & LRU & n/a \\
            \hline
            \revhpca{PACMAN~\cite{pacman}} & \centering LRU & \centering \revhpca{PACMAN} & \revhpca{LRU} & \revhpca{n/a} \\
            \hline
            \rev{SRRIP (L2C)~\cite{srrip}} & \centering \rev{SRRIP} & \centering \rev{SHiP} & \rev{LRU} & \rev{n/a} \\
            \hline
            \revhpca{DRRIP~\cite{srrip}} & \centering LRU & \centering \revhpca{DRRIP} & \revhpca{LRU} & \revhpca{n/a} \\
            \hline
            SHiP++~\cite{Young2017SHiP} & \centering LRU & \centering SHiP++ & LRU & n/a \\
            % \hline
            % TDRRIP~\cite{9804592} & \centering TDRRIP & \centering TSHiP & LRU & n/a \\
            \hline
            Mockingjay~\cite{9773195} & \centering LRU & \centering Mockingjay & LRU & n/a \\
            \hline
            \rowcolor{gray!45} \bcppr (Sec.~\ref{subsec:l2c_bcppr}) & \centering \bcppr & \centering SHiP & LRU & n/a \\
            
            \hline
            \hline
            \rowcolor{gray!45} tPB (Sec.~\ref{subsec:translation_prefetch_buffer}) & \centering LRU & \centering SHiP & LRU & LRU \\
            \hline
            tPB + CHiRP & \centering LRU & \centering \rev{SHiP} & CHiRP & LRU \\
            \hline
            tPB + CLIP & \centering CLIP & \centering SHiP & LRU & LRU \\
            \hline
            tPB + EMISSARY & \centering EMISSARY & \centering SHiP & LRU & LRU \\
            \hline
            tPB + PACIPV & \centering LRU & \centering PACIPV & LRU & LRU \\
            \hline
            \revhpca{tPB + PACMAN} & \centering LRU & \centering \revhpca{PACMAN} & \revhpca{LRU} & \revhpca{LRU} \\
            \hline
            \rev{tPB + SRRIP (L2C)~\cite{srrip}} & \centering \rev{SRRIP} & \centering \rev{SHiP} & \rev{LRU} & \rev{LRU} \\
            \hline
            \revhpca{tPB + DRRIP} & \centering LRU & \centering \revhpca{DRRIP} & \revhpca{LRU} & \revhpca{LRU} \\
            \hline
            tPB + SHiP++ & \centering LRU & \centering SHiP++ & LRU & LRU \\
            % \hline
            % tPB + TDRRIP & \centering TDRRIP & \centering TSHiP & LRU & LRU \\
            \hline
            tPB + Mockingjay & \centering LRU & \centering Mockingjay & LRU & LRU \\

            % \hline
            % \hline
            % CHiRP + Mockingjay & \centering LRU & \centering Mockingjay & CHiRP & n/a \\
            % \hline
            % CHiRP + EMISSARY & \centering EMISSARY & \centering SHiP & CHiRP & n/a \\
            % \hline
            % Morrigan + Mockingjay & \centering LRU & \centering Mockingjay & Morrigan & n/a \\
            % \hline
            % Morrigan + EMISSARY & \centering EMISSARY & \centering SHiP & Morrigan & n/a \\
            \hline
            
            \hline
            \hline
            \rowcolor{gray!45} \framework (tPB + \bcppr) & \centering \bcppr & \centering SHiP & LRU & LRU \\
            
            \hline
        \end{tabular}
    }
    % \vspace{-0.1cm}
    \caption{\rev{List and composition of all simulated designs.} % \gv{ADD SRRIP} 
    }
    \vspace{-0.4cm}
    \label{table:simulated_designs}
\end{table}

{\it Single-Core Workloads.} 
% Our set of single-core workloads consists of contemporary server workloads with large code footprints provided by Qualcomm for the CVP-1~\cite{cvp1} and IPC-1~\cite{ipc1} competitions, which have been used in recent literature~\cite{chirp, atp_sbfp, itpxptp}. We also use established server  workloads (NodeApp, PHPWiki, TPCC, Twitter, Wikipedia, Kafka, Spring, Tomcat, Chirper, HTTP), similar to previous work~\cite{schall2024llbp, thermometer}. Our evaluation considers only those server workloads with at least 0.5 instruction sTLB misses per kilo instructions. 
% Overall, our single-core evaluation considers 105 server workloads. Following a warm-up phase comprising 50 million instructions, our simulations execute an additional 100 million instructions to gather experimental results, similar to prior work~\cite{chirp,itpxptp}. 
%Our single-core evaluation uses 
We use a set of server workloads with large code footprints, including workloads provided by Qualcomm~\cite{cvp1,ipc1} and other established server workloads (NodeApp, PHPWiki, TPCC, Twitter, Wikipedia, Kafka, Spring, Tomcat, Chirper, HTTP),  used in recent literature~\cite{chirp, atp_sbfp, 10.1145/3669940.3707247, schall2024llbp,thermometer}. We consider only workloads exhibiting at least 0.5 instruction sTLB MPKI, resulting in a total of 105 single-core server workloads. After a warm-up phase of 50M instructions, simulations execute 100M instructions to collect experimental results~\rev{\cite{chirp,10.1145/3669940.3707247}}. %We only consider workloads with at least 0.5 instruction sTLB MPKI, yielding 105 single-core workloads. After a 50M-instruction warm-up, we simulate 100M instructions for evaluation~\rev{\cite{chirp,10.1145/3669940.3707247}}.

{\it Multi-Core Workloads.} We create both homogeneous and heterogeneous 4-core mixes using the single-core server workloads~\cite{bera_hermes,10476485,hawkeye,glider,multiperspective,jamet2024practically}. For the homogeneous mixes, we run four instances of each workload, one per core. For the heterogeneous mixes, we randomly combine four single-core workloads. In total, we consider 60 homogeneous and 100 heterogeneous mixes. 
The multi-core experiments use the same warm-up %(50M instructions) 
and simulation %(100M instructions) 
lengths as the single-core evalaution, with each workload running on its own core until at least one completes both phases~\cite{7095786}. We report the weighted speedup normalized to the baseline to avoid performance overestimation due to high-IPC threads~\cite{jimenez_multiperspective_2017,glider,psa}. For each single-core workload, we compute its IPC in a multi-core scenario shared with the other co-running single-core workloads ($IPC_{shared}$), and its IPC running alone on the same system ($IPC_{single}$). We then compute the weighted IPC of the mix as the weighted sum of $IPC_{shared}/IPC_{single}$ for all the benchmarks in the mix and we normalize this weighted IPC with the weighted IPC of the baseline.

\rev{{\it SMT Evaluation.} To evaluate \framework under SMT co-location, we extend ChampSim with SMT support. We construct 75 randomly selected workload pairs from the pool of single-core workloads to capture a diverse range of interference scenarios. For consistency, we use the same warmup and simulation lengths as in the single-core experiments. \autoref{smt_evaluation} presents the SMT evaluation.}

{\it Evaluated Policies.} \revhpca{We evaluate \rev{ten} state-of-the-art cache and TLB management policies: CHiRP~\cite{chirp}, Morrigan~\cite{morrigan}, CLIP~\cite{clip}, EMISSARY~\cite{emissary}, PACIPV~\cite{paciv}, PACMAN~\cite{pacman}, \rev{SRRIP~\cite{srrip},} DRRIP~\cite{srrip}, SHIP++~\cite{Young2017SHiP}, and Mockingjay~\cite{9773195}. Table~\ref{table:simulated_designs} lists these policies and the cache where they are applied following the standard practices. We also evaluate the \bcppr replacement (\autoref{subsec:l2c_bcppr}) in isolation as well as combining tPB (\autoref{subsec:translation_prefetch_buffer}) with all state-of-the-art policies; we exclude tPB+Morrigan due to poor performance.}
%(\autoref{subsec:translation_prefetch_buffer}), with CHiRP~\cite{chirp},  CLIP~\cite{clip}, EMISSARY~\cite{emissary}, PACIPV~\cite{paciv}, PACMAN~\cite{pacman}, \rev{SRRIP~\cite{srrip},} DRRIP~\cite{srrip}, SHIP++~\cite{Young2017SHiP}, and Mockingjay~\cite{9773195}. We observe that combining tPB with Morrigan~\cite{morrigan} produces poor performance and we omit this scenario in \autoref{sec:evaluation}.} %Overall, for each L1I prefetcher we consider \rev{22} scenarios (Table~\ref{table:simulated_designs})}.

    \section{Evaluation}
\label{sec:evaluation}

\begin{figure}
    \centering
    \includegraphics[width=\columnwidth]{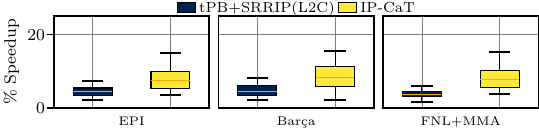}
    \vspace{-0.6cm}
    \caption{\rev{Performance of \framework and tPB+SRRIP (best-performing policy of Fig.~\ref{fig:evaluation:all_speedups}) across 788 workloads without the MPKI selection of \autoref{sec:methodology}.}}
    \vspace{-0.4cm}
    \label{fig:evaluation:single_core:all_apps}
\end{figure}

\rev{To demonstrate the effectiveness of \framework, we evaluate three state-of-the-art L1I prefetchers: EPI~\cite{epi_isca}, Barça~\cite{gratz2020barca}, and FNL+MMA~\cite{fnlmma}. All are configured to prefetch across page boundaries, as \autoref{subsec:Instruction_prefetchers} shows that it provides significant performance gains.} \rev{Across the studies server workloads, these prefetchers exhibit both high accuracy and coverage: EPI achieves 74.1\% accuracy and 85.8\% coverage, FNL+MMA 72.9\% and 85.0\%, and Barça 67.7\% and 83.7\%. Evaluating with such strong L1I prefetchers avoids overstating \framework’s benefits, which could otherwise be inflated by low-accuracy or low-coverage prefetch engines.}

\subsection{Single-Core Performance Evaluation}
\label{subsec:overall}

This section compares the single-core performance of all schemes in \autoref{table:simulated_designs}. \autoref{fig:evaluation:all_speedups} reports results using EPI, Barça, and FNL+MMA as L1I prefetchers. The x-axis lists the evaluated schemes and the y-axis shows the speedups over the baseline (\autoref{table:champsim}). Each scheme is represented with a box plot, showing the distribution of speedups across the 105 server workloads, with a red bar indicating the geometric mean.

Regarding the \bcppr component of \framework, we observe that it delivers \rev{2.9\%, 4.8\%, and 5.0\%} geomean speedups across the EPI, Barça, and \shp{FNL+MMA} prefetchers, outperforming the state-of-the-art cache and replacement policies. We observe such behavior because \bcppr tailors the insertion, promotion, and eviction policies of prefetched code lines in L2C to the underlying execution phase, as explained in \autoref{subsec:l2c_bcppr}.

\framework (\tpb combined with \bcppr) improves the performance of EPI, Barça, and \shp{FNL+MMA} by \rev{6.1\%, 8.3\% and 7.9\%} across the  EPI, Barça, and \shp{FNL+MMA} prefetchers, outperforming CHiRP, Morrigan, CLIP, EMISSARY, PACIPV, PACMAN, DRRIP, SHiP++, and \shp{Mockingjay} across all L1I prefetchers even when these schemes are combined with tPB. \framework's benefits stem from the fact that \bcppr  optimizes the management of prefetched code lines in L2C while tPB serves a significant number of demand sTLB misses while improving the timeliness of L1I prefetching, as \autoref{subsec:evaluation:perf_analysis} shows. 

\rev{\emph{Non TLB Intensive Workloads.} To highlight that \framework does not harm the performance of non TLB intensive applications, we evaluate all server workloads provided by Qualcomm~\cite{cvp1,ipc1} without the sTLB MPKI selection of \autoref{sec:methodology}. For this study, we compare \framework to the best-performing policy of \autoref{fig:evaluation:all_speedups}: tPB+SRRIP. \autoref{fig:evaluation:single_core:all_apps} shows the performance of tPB+SRRIP and \framework across all L1I prefetchers. For EPI, \framework outperforms tPB+SRRIP by 2.9\% in geomean, indicating the benefits of our proposal. The trends for FNL+MMA and Barça are similar.}

\begin{figure}
    \centering
    \includegraphics[width=1\columnwidth]{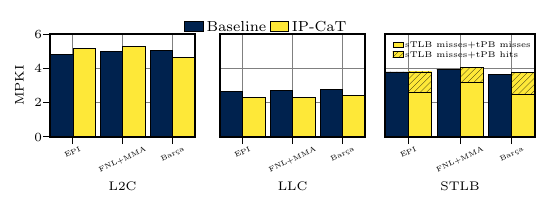}
    \vspace{-0.6cm}
    \caption{
    \revhpca{L2C, LLC, and STLB MPKIs across all prefetchers. % \gv{it would be easier to name the legend: sTLB misses + tPB misses and sTLB misses + tPB hits}
    % \alex{Once we agree on the numbers I'll also shrink these plots.}
    }
    }
    \vspace{-0.4cm}
    \label{fig:evaluation:l2c_llc_stlb_mpki}
\end{figure}

\begin{figure}
   \centering
   \includegraphics[width=1\columnwidth]{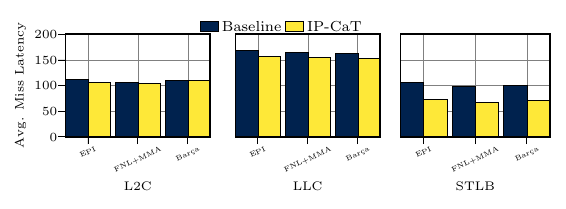}
   \vspace{-0.6cm}
   \caption{\rev{L2C, LLC, and STLB average miss latency.} 
   }
   \vspace{-0.4cm}
   \label{fig:evaluation:l2c_llc_stlb_avg_miss_latency}
\end{figure}

\subsubsection{Performance Analysis}
\label{subsec:evaluation:perf_analysis}

% \gv{this section needs better writing flow.}
To explain \framework's superior performance, \autoref{fig:evaluation:l2c_llc_stlb_mpki} and \autoref{fig:evaluation:l2c_llc_stlb_avg_miss_latency} show its impact on MPKI and average miss latency, respectively, for the L2C, the LLC \rev{ (excluding misses due to address translations)}, and the sTLB.
%show the impact of \framework on the MPKI and average miss latency concerning L2C and LLC, \rev{excluding misses due to address translations}, and sTLB.
% Figures~\ref{fig:evaluation:l2c_llc_stlb_mpki} and~\ref{fig:evaluation:l2c_llc_stlb_avg_miss_latency} also show measurements for the baseline system of \autoref{table:champsim}, shown under the {\it Baseline} category.\gv{referring to the baseline is not needed.}
We observe that the impact of \framework on the cache and TLB hierarchy is twofold:  
i) \bcppr reduces the number of LLC misses for all L1I prefetchers, while slightly increasing the L2C misses for EPI and FNL+MMA \rev{and simultaneously achieving a significant reduction in the average miss latency for L2C, LLC, and sTLB}. 
% The MPKI reduction primarily arises from the decrease in instruction misses in these levels of the memory hierarchy\gv{UNCLEAR, NOT TRUE}. 
The MPKI reduction primarily arises from better management of lines brought in the L2C by L1i prefetch requests and a decrease in traffic due to page walks. 
At L2C, \bcppr increases instruction MPKI by 6.5\%, 5.1\%, and -8.1\% for EPI, Barça, and FNL+MMA, respectively \rev{while reducing the average miss latency by 3.8\%, 2.3\%, and -0.5\%, respectively};  
ii) \tpb substantially reduces both sTLB MPKI and the average miss latency by more effectively managing address translation entries brought in by L1I page-crossing prefetches.
% \revhpca{While the sTLB MPKI of \framework is very similar as the baseline, the average sTLB miss latency is reduced by 30.2\%, 30.8\%, and 26.4\%  for EPI, Barça and FNL+MMA, respectively, since a significant portion of sTLB misses are served by \tpb and thus do not require triggering page walks.
% This page walk count reduction also decreases the pressure that sTLB misses put on the cache hierarchy, and therefore L2C and LLC average miss latency also experience reductions, as \autoref{fig:evaluation:l2c_llc_stlb_avg_miss_latency} shows.
% These two effects explain the performance improvements achieved by \framework (\autoref{fig:evaluation:all_speedups}).}
%While the sTLB MPKI of \framework is very similar as the baseline, 
%First, 
We observe that the sTLB MPKI, defined as the number of accesses that miss in both the sTLB and the \tpb, is reduced by 31.6\%, 18.2\%, and 32.3\% for EPI, FNL+MMA, and Barça, respectively. This reduction occurs because a substantial fraction of demand sTLB misses are served by tPB. %Second, we observe that the average sTLB miss latency is reduced by 30.2\%, 30.8\%, and 26.4\%  for EPI, Barça and FNL+MMA, respectively, as misses saved by \tpb no longer require triggering page walks. 
This reduction in sTLB MPKI also decreases the pressure that sTLB misses put on the cache hierarchy, therefore L2C and LLC average miss latencies experience reductions (\autoref{fig:evaluation:l2c_llc_stlb_avg_miss_latency}).

\begin{figure}[ht]
    \centering
    \includegraphics[width=1\columnwidth]{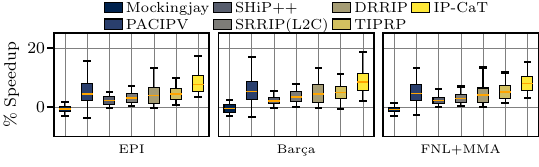}
    \vspace{-0.6cm}
    \caption{
    \rev{Comparison against L2C variants of the state-of-the-art replacement policies. % \gv{fix legend}
    }
    }
    \vspace{-0.2cm}
    \label{fig:l2c_variants_comp}
\end{figure}

\subsubsection{\revhpca{Comparison with L2C Variants of Prior Policies}}
\label{subsubsec:evaluation:single_core:l2c_variants_comp}

This section compares \bcppr and \framework with the state-of-the-art policies of \autoref{table:simulated_designs} which are originally designed for the LLC (Mockingjay, PACIPV, SHiP++, \rev{SRRIP, and} DRRIP), now applied to L2C. We exclude PACMAN from this study due to its inferior performance. \autoref{fig:l2c_variants_comp} presents the performance comparison, showing that \framework provides higher speedups than the competing policies owing to the \bcppr and tPB schemes that optimize the management of prefetched code lines and code translations in L2C and sTLB, respectively. For example, when EPI is used, \framework outperforms Mockingjay, PACIPV, SHiP++, \rev{SRRIP, and} DRRIP by \rev{8.0\%, 3.0\%, 5.3\%, 4.4\%, and 3.6\%}, respectively.

\begin{figure}
    \centering
    \includegraphics[width=1\columnwidth]{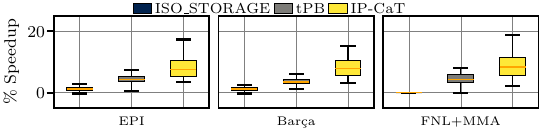}
    \vspace{-0.6cm}
    \caption{\rev{Comparing \framework against an augmented sTLB.}}
    \vspace{-0.4cm}
    \label{fig:ip_cat_vs_iso_stlb}
\end{figure}

\subsubsection{ISO Storage Comparison}
\label{subsec:evaluation:iso_storage_comp}

\autoref{fig:ip_cat_vs_iso_stlb} compares \revhpca{tPB} and \framework with a scenario that augments the sTLB with \framework's storage overhead. \rev{This is done by adding one way to the sTLB, providing 128 extra sTLB entries as opposed to tPB’s 64 entries (\autoref{subsec:putting_it_together}).} The results of \autoref{fig:ip_cat_vs_iso_stlb} show that \tpb and \framework outperforms this  \texttt{ISO\_Storage} scenario across all considered L1I prefetchers. 

\begin{table}[ht]
    \centering
    \setlength\tabcolsep{10pt} 
    \renewcommand{\arraystretch}{0.5} 
    \resizebox{1\columnwidth}{!}{
        \begin{tabular}{|l|c|c|c|c|c|}
            \hline
            \rowcolor{gray!30} \textbf{Table Groups} & \centering \textbf{T0} & \centering \textbf{T1-T2} & \textbf{T3-T10} & \textbf{T11-T15} & \textbf{Total} \\
            \hline
            \hline
            \textbf{Entries} & 4K & 4K & 4K & 2K & 14K \\
            \hline
            \textbf{Tag bits} & 0 & 9 & 13 & 15 & \\
            \hline
            \textbf{U bits} & 0 & 1 & 1 & 1 & \\
            \hline
            \textbf{bits per entry} & 27 & 37 & 41 & 43 & \\
            \hline
            \textbf{Storage (KBits)} & 108 & 148 & 164 & 86 & 506 \\
            \hline
        \end{tabular}
    }
    \vspace{-0.1cm}
    \caption{Configuration of the ITTAGE's prediction tables.
    }
    \vspace{-0.6cm}
    \label{table:ittage_config}
\end{table}

\begin{figure}[ht]
    \centering
    \includegraphics[width=1\columnwidth]{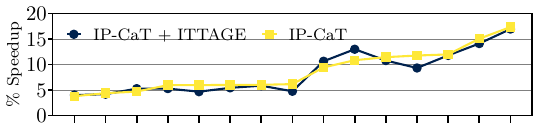}
    \vspace{-0.6cm}
    \caption{\shp{Evaluation of \framework considering a baseline with TAGE-SC-L, and a scenario using TAGE-SC-L with ITTAGE.} 
    }
    \vspace{-0.4cm}
    \label{fig:evaluation:ittage_comp}
\end{figure}

\subsection{\shp{Impact of Indirect Branch Target Prediction}}
\label{subsec:evaluation:impact_indirect_target_prediction}

Our baseline uses TAGE-SC-L as conditional branch predictor. This section evaluates \framework with and without the state-of-the-art indirect branch predictor ITTAGE~\cite{seznec:hal-00639041} to quantify its impact on the performance of our proposal. We consider two configurations: i) the branch prediction unit (BPU) of the baseline uses only TAGE-SC-L, as in prior sections, and ii) the BPU of the baseline uses both TAGE-SC-L and ITTAGE. The ITTAGE configuration is detailed in \autoref{table:ittage_config}. 

% \autoref{fig:evaluation:ittage_comp} presents the performance of \framework across 15 representative server workloads under both configurations. We observe that \framework delivers very similar speedups over both baselines, with a maximum speedup variation of 2.4\%. All the competing policies of \autoref{fig:evaluation:all_speedups} show a similar behavior with and without ITTAGE, which means that considering a baseline with an indirect branch prediction does not reduce \framework benefits.
\autoref{fig:evaluation:ittage_comp} shows the performance of \framework across 15 representative server workloads under both configurations. The results are nearly identical, with a maximum IPC variation of 2.4\%. The competing policies of \autoref{table:simulated_designs} exhibit similar trends in both setups, thus incorporating ITTAGE in the baseline does not reduce the benefits of \framework.

% \subsection{Performance Contribution of \framework Components}
\subsection{Ablation Study of \framework Components}
\label{subsec:evaluation:breakdown_epic_performance}

\begin{figure}%[h]
    \centering
    \includegraphics[width=1\columnwidth]{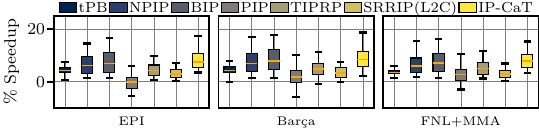}
    \vspace{-0.6cm}
    \caption{
    \rev{Performance breakdown of \framework.}}
    \vspace{-0.2cm}
    \label{fig:breakdown_l2c_repl}
    % \vspace{-0.4cm}
\end{figure}

\begin{figure}%[h]
    \centering
    \includegraphics[width=1\columnwidth]{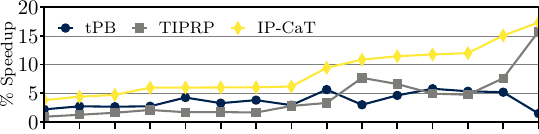}
    \vspace{-0.6cm}
    \caption{\rev{Comparison between \tpb, \bcppr, and \framework considering the EPI prefetcher across representative workloads.} % \gv{this can be smaller in height similar to figure 14}
    }
    \vspace{-0.2cm}
    \label{fig:evaluation:tpb_vs_tiprp_line_plot}
\end{figure}

This section quantifies the contribution of each \framework component. We evaluate: i) \tpb; ii) \npcp, iii) \bip, iv) \pcp, and v) \bcppr as standalone L2C replacement policies; vi) SRRIP since \npcp, \bip, and \pcp are based on it, and vii) \framework, which combines \tpb and \bcppr. \autoref{fig:breakdown_l2c_repl} reports the results. With the EPI prefetcher, \tpb, \npcp, \bip, \pcp, \bcppr, SRRIP, and \framework achieve 2.9\%, 4.8\%, 5.5\%, -1.5\%, 2.9\%, 1.7\%, and 6.1\% geomean speedups, respectively. Notably, \framework outperforms the sum of its components (\tpb + \bcppr = 5.8\%), reaching 6.1\%. This gain comes from the synergy between \tpb and \bcppr: since page walks access the L2C, \tpb reduces the number of page walks, lowering L2C contention and increasing the effectiveness of \bcppr. \autoref{fig:evaluation:tpb_vs_tiprp_line_plot} corroborates this across 15 representative server workloads (same as the ones used in \autoref{fig:evaluation:ittage_comp}). In particular, \tpb reduces L2C MPKI by 3.6\% due to the reduction in page walks, enabling larger gains for \bcppr. We observe similar trends with Barça and FNL+MMA prefetchers.

\begin{figure}[ht]
    \centering
    \vspace{-0.2cm}
    \includegraphics[width=1\columnwidth]{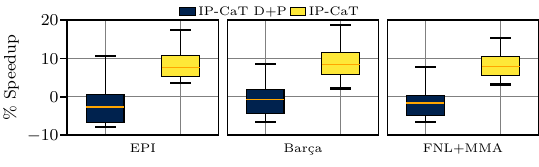}
    \vspace{-0.6cm}
    \caption{
    \rev{Comparison to a variation of \framework applying \bcppr to both demand and prefetch code lines in L2C (\framework D+P).}
    }
    \vspace{-0.2cm}
    \label{fig:ipcat_d_p_comp}
\end{figure}

\subsubsection{Applying \bcppr to Demand Instruction Accesses} 
\label{sec:ipcat_d_p_comp}

\autoref{fig:ipcat_d_p_comp} compares \framework to a variation of \framework that applies the \bcppr replacement policy not only to lines fetched by L1I prefetches but also to demand instruction accesses; we refer to this scheme as \framework D+P. %\rev{This scheme is achieve by not considering whether an L2C line was inserted by an L1i prefetch request when performing the promotions, evictions, and policy updates at the L2C. Therefore, all lines inserted by an L1i prefetch request are treated equally.} 
\autoref{fig:ipcat_d_p_comp} shows that, when considering the EPI prefetcher, \framework outperforms (in geomean) \framework D+P by \revhpca{\rev{10.1\%}}. This study indicates that applying \bcppr to both prefetch and demand instruction requests harms performance since
cache lines fetched by demand instruction accesses show different reuse patterns compared to lines fetched by instruction prefetches. 

\begin{figure}
    \centering
    \includegraphics[width=\linewidth]{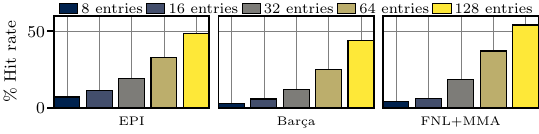}
    \vspace{-0.6cm}
    \caption{\rev{Sensitivity to \tpb size, assuming fully-associative \tpb.} 
    }
    \vspace{-0.2cm}
    \label{fig:evaluation:tpb_size_sensitivity_analysis}
\end{figure}

\begin{figure}
    \centering
    \includegraphics[width=\linewidth]{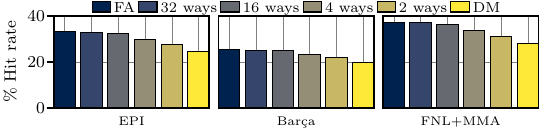}
    \vspace{-0.6cm}
    \caption{\rev{Sensitivity analysis on the associativity of a 64 entries \tpb, varying from fully-associative to direct-mapped. % \gv{height can be smaller for fig 18 and 19 and 20}
    } 
    }
    \vspace{-0.2cm}
    \label{fig:evaluation:tpb_associativity_sensitivity_analysis}
\end{figure}

\subsection{\rev{Sensitivity to \tpb Size and Organization}}
\label{subsec:evaluation:sensitivity_to_tpb_size}

\rev{\autoref{fig:evaluation:tpb_size_sensitivity_analysis} presents a sensitivity analysis of the \tpb size in terms of hit rate. In this analysis the \tpb is a fully associative standalone structure with capacities ranging from 8 to 128 entries. For the EPI prefetcher the \tpb hit rate monotonically increases from 3.1\% to 48.3\% when its size grows from 8 to 128 entries. We observe similar trends for the other prefetchers. Based on this trade-off, we select a 64-entry \tpb as it represents a practical design point between coverage and hardware complexity; unless otherwise stated, all results presented in the paper use this 64-entry \tpb.}
%\mc{Remove 18?}

% \gv{plot2: pick one size and do sensitivity analysis on associativity}

\rev{
%For the selected 64-entry configuration, we next analyze the impact of associativity on tPB effectiveness. 
\autoref{fig:evaluation:tpb_associativity_sensitivity_analysis} presents a sensitivity study of tPB's hit rate as we vary its organization from fully associative to direct-mapped, while keeping the total number of entries fixed at 64.}
\rev{For the EPI L1i prefetcher, the tPB hit rate decreases from 37.2\% to 28.0\% when moving from a fully associative to a direct-mapped organization. Notably, we observe that the difference in hit rate between the fully-associative, 32-way, and 16-way organizations of \tpb is rather small. Although in this paper we consider the fully-associative design as our primary \tpb design, the design comprising 4 sets and 16 ways performs similarly and may constitute more practical configuration.} 
% \gv{which remainder of the analysis? All prior sections assume a fully associative tpb right? we should say that the diff is very small and for our experiments we consider a fully associative tpb. However, someone could go with the 16w designs since it's the one closer to fully assoc.}

%\rev{\autoref{fig:evaluation:tpb_augmented_stlb_12w} compares the performance improvement of a fully-associative standalone \tpb to two designs augmenting the sTLB by 4 and 8 ways, respectively. We observe that these augmented sTLB designs behave very similarly to the standalone \tpb. When considering the EPI L1i prefetcher, the hit rate improvement ranges from 25.6\% to 41.6\% geometric mean speedup over the baseline.}
%\mc{add the sTLB hit rate when using a standalone tPB? Remove paragraph?}
% \mc{Text to be adapted to hits/misses. % Shouldn't this text appear in V.C?
% }

% \subsection{\rev{\framework Performance When Integrating \tpb in sTLB}}
\subsection{Integrating \tpb in sTLB}
\label{subsec:evaluation:integrating_tpb_in_stlb}

\begin{figure}
    \centering
    \includegraphics[width=1\columnwidth]{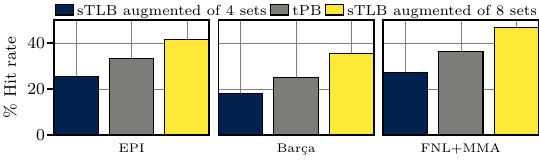}
    \vspace{-0.6cm}
    \caption{\rev{Comparing \framework with a 64-entry fully-associative \tpb against augmenting sTLB by 4 and 8 sets of 12 ways.} 
    }
    \vspace{-0.2cm}
    \label{fig:evaluation:tpb_augmented_stlb_12w}
\end{figure}

\rev{
This section evaluates two variations of \framework involving a \tpb with the same number of ways as the sTLB. 
\autoref{sss:integration} describes how a \tpb with the same number of ways as the sTLB can be seamlessly integrated into it.
Specifically, we consider two designs which augment the sTLB with 4 and 8 additional sets of 12 ways each, respectively.
For completeness, we also show the fully-associative design which decoupled \tpb from sTLB, presented in all previous sections. \autoref{fig:evaluation:tpb_augmented_stlb_12w} shows the hit rates of the scenarios which integrate \tpb in the sTLB as well as the hit rate of the standalone fully-associative \tpb. The latter exhibits a 36.2\% hit rate whereas the two sections of the augmented sTLB exhibit 25.6\% and 41.6\% hit rate when augmenting the sTLB with 4 and 8 sets dedicated to \tpb, respectively.
These differences in terms of \tpb hit rates across the three designs do not translate into significant \framework performance differences.} 
%\mc{delete fig 20?}

\begin{figure}
    \centering
    \includegraphics[width=1\columnwidth]{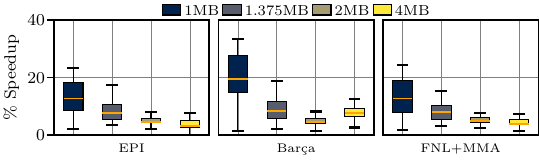}
    \vspace{-0.6cm}
    \caption{\shp{Sensitivity of \framework's performance to the LLC size.} 
    }
    \vspace{-0.4cm}
    \label{fig:evaluation:sensitivity_analysis_llc}
\end{figure}

\subsection{\shp{Sensitivity to LLC size}}
\label{subsec:evaluation:sensitivity_analysis}

\shp{\autoref{fig:evaluation:sensitivity_analysis_llc} presents a sensitivity analysis of \framework performance as the LLC capacity is varied from 1MB to 4MB. With a 1MB LLC, \framework achieves a geometric mean speedup of 12.7\% for the EPI prefetcher, while the speedup reduces to 2.6\% with a 4MB LLC. The main takeaway is that even with larger LLCs ($e.g.$, 4MB), \framework continues to provide significant performance improvements. Similar trends are observed for the FNL+MMA and Barça.}

\shp{An additional observation from \autoref{fig:evaluation:sensitivity_analysis_llc} is that the performance gains of \framework gradually decrease as the LLC size increases across all evaluated L1I prefetchers. This happens because large LLCs can capture a greater portion of application working sets, thereby improving cache locality and reducing miss rates, which in turn makes the L2C replacement policy less critical for performance. Consequently, the relative contribution of \bcppr on \framework speedups becomes less pronounced, resulting in smaller performance improvements.}

\begin{figure}
    \centering
    \begin{subfigure}{1\columnwidth}
        \includegraphics[width=\linewidth]{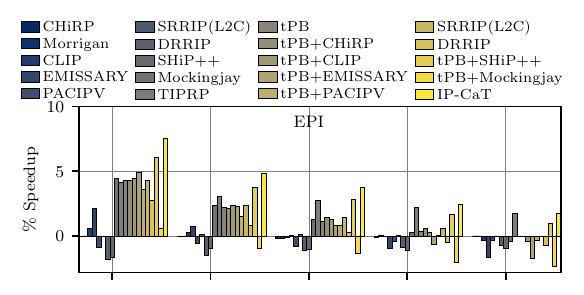}
        \label{subfig:evaluation:epi_multi_page_size_speedups}
    \end{subfigure}
    \vspace{-1.2cm}
    \\
    \begin{subfigure}{\linewidth}
        \includegraphics[width=\linewidth]{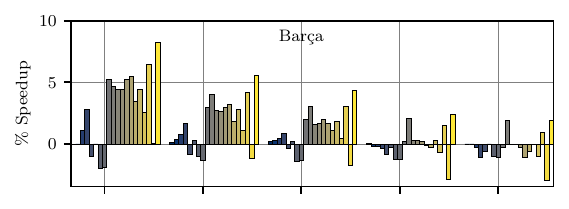}
        \label{subfig:evaluation:barca_multi_page_size_speedups}
    \end{subfigure}
    \vspace{-1.2cm}
    \\
    \begin{subfigure}{\linewidth}
        \includegraphics[width=\linewidth]{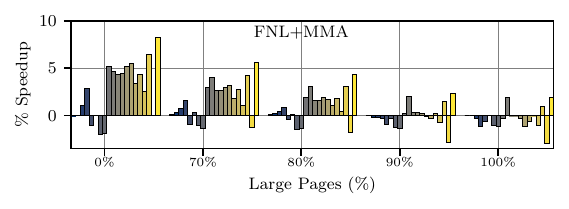}
        \vspace{-0.4cm}
        \label{subfig:evaluation:fnl_mma_multi_page_size_speedups}
    \end{subfigure}
    \vspace{-0.6cm}
    \caption{\rev{Evaluation on multiple page sizes.}}
    % \vspace{-0.1cm}
    \label{fig:evaluation:all_multi_page_speedups}
\end{figure}

% \subsection{Evaluation with Multiple Page Sizes}
\subsection{Multiple Page Sizes}
\label{subsec:evaluation:multi_page_size_performance_analysis}

\rev{\autoref{fig:evaluation:all_multi_page_speedups} shows the performance improvement of all considered scenarios (\autoref{table:simulated_designs}), excluding PACMAN due to inferior performance, when the baseline uses both 4KB and 2MB pages, as explained in \autoref{sec:methodology}.} The top, medium, and bottom plots show results for EPI, Barça, and FNL+MMA, respectively.
The x-axis reports the proportion of the memory footprint mapped in large pages as compared to small pages (\textit{e.g.}, 5\% refers to a scenario where 5\% of the memory footprint is mapped in 2MB pages; the remaining 95\% is mapped to 4KB pages). 
The y-axis shows the geomean speedups over the baseline for each multi-size page scenario. 

We observe that \framework consistently outperforms all state-of-the-art approaches for all multi-page size scenarios and L1I prefetchers. For example, with the EPI prefetcher, the geomean speedup of \framework goes from \rev{7.5\% to 1.8\%} as the proportion of the memory footprint mapped to 2MB pages increases from 0\% to 100\%.  The best state-of-the-art scheme moves from \rev{4.5\% to -0.4\%} as the footprint mapped into 2MB pages goes from 0\% to 100\%. \tpb alone does not provide any benefit when the entire memory footprint is mapped in 2MB pages since the number of page-cross prefetch requests missing in the sTLB is minimal in this scenario. Overall, the benefits of \framework (and all competing approaches) diminish as a larger fraction of code and data is mapped to 2 MB pages, since using 2MB pages reduces STLB misses. Nevertheless, even when the entire code and data footprint 2MB pages, \framework still achieves a non-negligible 1.8\% speedup over the baseline.

\begin{figure}
    \centering
    \vspace{-0.4cm}
    \includegraphics[width=1\columnwidth]{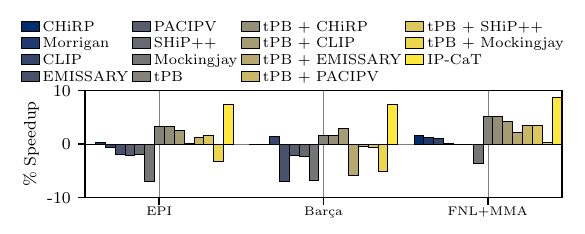}
    \vspace{-0.6cm}
    \caption{Performance evaluation in 4-core context. % \gv{fix legend}
    }
    \vspace{-0.6cm}
    \label{fig:evaluation:multi_core_performance}
\end{figure}

\subsection{Multi-Core Evaluation}
\label{subsec:multi_core_performance_analysis}

This section quantifies the performance of \framework in multi-core contexts. \autoref{fig:evaluation:multi_core_performance} presents the speedups of \framework and the other schemes listed in \autoref{table:simulated_designs} over a 4-core baseline across 160 workload mixes, presented in \autoref{sec:methodology}. \autoref{fig:evaluation:multi_core_performance} reveals that \framework outperforms all competing schemes across all considered L1I prefetchers. For the EPI prefetcher, \framework outperforms CHiRP, Morrigan, CLIP, EMISSARY, PACIPV, SHiP++, and \shp{Mockingjay} by 7.2\%, 6.8\%, 7.8\%, 9.1\%, 9.3\%, 9.0\%, and 14.2\%, respectively. Although combining \tpb with the state-of-the-art schemes improves performance, combining \tpb with \bcppr, $i.e.$ \framework, delivers the best performance in the multi-core context.

\begin{figure}
    \centering
    \includegraphics[width=1\columnwidth]{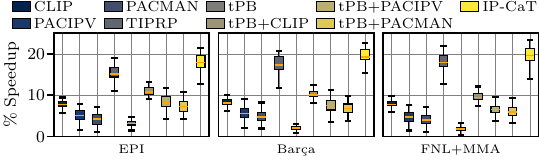}
    \vspace{-0.6cm}
    \caption{\rev{Performance evaluation using SMT workloads. % \gv{legend can be improved.}
    } % \gv{bigger legend + a bit up}
    }
    \vspace{-0.4cm}
    \label{fig:evaluation:smt_workloads}
\end{figure}

\subsection{\rev{SMT Evaluation}}
\label{smt_evaluation}

\rev{\autoref{fig:evaluation:smt_workloads} quantifies the performance of \framework and the other prefetch-aware schemes considering 75 SMT workloads, presented in \autoref{sec:methodology}. We observe that \framework outperforms all competing schemes across all considered prefetchers. The results show similar trends as in the single-thread evaluation, but with larger absolute speedups due to increased contention for structures such as the sTLB and L2C. For example, when considering the EPI, \framework outperforms CLIP, PACIPV, and PACMAN by 7.1\%, 9.3\%, and 10.3\%, respectively.}
    \section{Related Work}
\label{sec:related_work}

% To the best of our knowledge this paper proposes  the first hardware approach for effective L1I cross-page prefetching, \framework. By reducing the address translation costs that undermine the efficacy of L1I prefetchers, \framework achieves substantial performance improvements for workloads with large instruction footprints.
% \framework does not pollute the L1I neither the TLB with inaccurate prefetch blocks while, at the same time, populates the tPB with instruction translations that enable effective L1I cross-
% page prefetching.
% There are previous works addressing similar topics:

% \emph{Instruction Prefetching.} The interest in L1I prefetching has grown due to the advent of emerging server and datacenter applications with large code footprints. State-of-the-art L1I prefetchers \cite{fnlmma, epi_isca, gratz2020barca, djolt, fdip, ipc1} provide significant performance gains. When these prefetchers cross page boundaries and generate page table walks, they become less timely, limiting  their potential for IPC gains. This work optimizes L1I page-cross prefetching by mitigating its translation costs while reducing L2C pollution by anticipating whether lines fetched in L2C by L1I page-cross prefetch requests will be reused or not.

\rev{\emph{Cache Management and Set Dueling.} \framework extends beyond the monolithic behavior of traditional set-dueling–based policies like BIP and DIP~\cite{10.1145/1250662.1250709}, as Section~\ref{subsubsec:dynamically_selecting_between_pcp_and_npcp} explains. While these approaches use a global counter and update it only on evictions, \bcppr employs a two-level decision tree in which node-level decisions are driven by saturating counters. This structure enables finer-grained adaptation of replacement decisions. In addition, \bcppr updates its control counters on both L2C hits and evictions, rather than only on evictions, allowing it to react more quickly to workload behavior. By also distinguishing between lines brought by L1I prefetches and other lines, the policy captures differences in line utility that traditional set-dueling policies overlook.} 

\revhpca{\emph{Page-Cross Prefetching.} 
Prior art~\cite{vavou25} reveals that page-crossing for data is seldom beneficial across different applications types and access patterns. Our work shows that when state-of-the-art L1I prefetchers cross page boundaries, the vast majority of the corresponding prefetches are accurate. 
This difference stems from the fact that instruction accesses typically follow sequential and loop-based control flow patterns, leading to highly predictable streams, while data access patterns are often hard to predict (\emph{e.g.}, pointer chasing)~\cite{makis1,makis2}. %{\color{red}Indeed, EPI's~\cite{epi_isca} accuracy on page-cross prefetching is $>$75\% while the accuracy of the state-of-the-art L1D prefetcher, Berti~\cite{berti}, is just$~$50\% when it comes to page-cross prefetching~\cite{vavou25}.}
}

% \emph{Page-Size Aware Prefetching.} Prior work \cite{psa} highlights that data prefetchers placed alongside PIPT caches (L2C, LLC) typically do not trigger prefetches that cross 4KB page boundaries for security reasons. The authors propose a scheme that makes lower-level cache prefetchers to safely cross 4KB page boundaries by exploiting the existence of larger page sizes. Our proposal is orthogonal to this work as we i) target instruction page-cross prefetching for the first-level instruction cache (L1I) which operates with virtual addresses, thus the L1I prefetcher can go beyond page boundaries without any security implication and ii) optimize sTLB and L2C management for L1I page-cross prefetch requests. 

\emph{Translation Prefetching.} 
%Prior art targets data sTLB prefetching~\cite{Kandiraju:2002,Pham:2015,Baer:1995,Kandiraju:2002,atp_sbfp} and instruction sTLB prefetching~\cite{morrigan}. 
Previously proposed TLB prefetchers~\cite{Kandiraju:2002,Pham:2015,Baer:1995,Kandiraju:2002,atp_sbfp,vavouliotis_free_tlb_prefetching,vavouliotis_vm_prefetching} typically use buffers to store the prefetched  translations to avoid sTLB pollution. % the sTLB content in case of inaccurate prefetching. 
The tPB component of \framework is orthogonal to sTLB prefetching since it is populated only by instruction translation requests originating from L1I page-cross prefetches and does not trigger additional sTLB prefetches. Our evaluation shows that \framework outperforms the state-of-the-art instruction sTLB prefetcher~\cite{morrigan}. % in a wide range of scenarios due to optimizing cache and TLB management for L1I prefetches. 

\emph{TLB Management.} POM-TLB~\cite{pom_tlb} uses a large die-stacked TLB to reduce number of page walks. Victima~\cite{victima} uses a part of L2C as L3 TLB to store evicted sTLB data entries. DVMT~\cite{Alam:2017} allows the application to define the page table format to reduce the page walk overhead. Elnawawy et al.~\cite{diligent_tlbs} pins in the sTLB highly used data PTEs. Our work is orthogonal to these approaches since tPB stores only translations fetched by L1I page-cross prefetches. % and \bcppr optimizes L2C management for line fetched by L1I prefetches.

\emph{Cooperative TLB and Cache Policies.} Chasapis et al.~\cite{10.1145/3669940.3707247} combines an sTLB replacement policy (iTP) that maximizes the number of instruction hits in the sTLB at the expense of increasing the number of data page walks with an L2C policy (xPTP) that accelerates data page walks \rev{by giving higher priority to data translations lines in L2C over the other line types}. \framework is orthogonal to \cite{10.1145/3669940.3707247} as it accelerates L1I page-cross prefetching by storing \rev{speculatively fetched, by the L1I prefetcher,} PTEs in tPB and optimizes the management of lines fetched in L2C by L1I prefetches without applying any specialized management for translation lines. Combining iTP and tPB at the sTLB and xPTP and \bcppr at L2C while using a smart selection scheme has great potential.

% \rev{\emph{Code Layout Optimizations.} Profile-guided code layout techniques such as BOLT~\cite{bolt} and Codestitcher~\cite{codestitcher} improve instruction locality by reordering functions and basic blocks to reduce instruction cache and TLB pressure. Compile-time approaches~\cite{10.5555/3049832.3049858,llinux} place hot functions in huge pages to reduce translation overheads. OS-level schemes leveraging superpages~\cite{10.555,10.1145} map small code regions into superpages via superpage promotion and page table sharing. Recency-based TLB Preloading~\cite{Saulsbury:2000} builds a recency stack of PTEs in the page table to derive prefetches based on past access patterns. These techniques primarily reduce instruction footprint or modify virtual-to-physical mappings through software or OS support. \framework is a synergistic microarchitectural scheme that requires no page table or software modifications and remains complementary to code layout optimizations, addressing translation and caching limitations that persist even after the application of code layout optimizations.}

\rev{\emph{Code Layout Optimizations.} Profile-guided techniques such as BOLT~\cite{bolt} and Codestitcher~\cite{codestitcher} improve instruction locality by reordering functions and basic blocks to reduce I-cache and TLB pressure. Compile-time approaches~\cite{10.5555/3049832.3049858,llinux} place hot code in huge pages, while OS-level schemes use superpages~\cite{10.555,10.1145} via promotion or page table sharing to reduce translation overhead. Recency-based TLB preloading~\cite{Saulsbury:2000} predicts future accesses from past PTE reuse. These methods primarily reduce code footprint or rely on software/OS changes. \framework is a microarchitectural solution that requires no changes to page tables or software and remains complementary to code layout optimizations, addressing translation and caching limitations that persist even after the application of code layout optimizations.}
    \section{Conclusions}
\label{sec:conclusions}

This work demonstrates that the address translation latency of L1I prefetches that cross page boundaries and the variable behavior of lines fetched in L2C by L1I prefetches undermines the benefits of modern L1I prefetchers. To address these limitations, this work proposes \emph{\frameworkfullname}, the first microarchitectural scheme to orchestrate TLB and cache management to maximize the benefits of L1I prefetching for applications with large code footprints. Our evaluation shows that \framework significantly enhances the performance of state-of-the-art L1I prefetchers and outperforms leading TLB and cache management policies across 105 single-core and 160 multi-core server workloads, with only 0.79KB of storage overhead.
% \end{spacing}

\iffalse
    \appendix
    
    \input{09_artifact_evaluation}
\fi
%%%%%%% -- PAPER CONTENT ENDS -- %%%%%%%

\section*{Acknowledgment}

Alexandre Valentin Jamet acknowledges his AI4S fellowship within the “Generación D” 
initiative by Red.es, Ministerio para la Transformación Digital y de la Función Pública, for 
talent attraction (C005/24-ED CV1), funded by NextGenerationEU through PRTR.
This work has received funding from ‘Future of Computing, a Barcelona Supercomputing Center and IBM initiative’ (2023). It has  been partially supported by the project PID2023-146511NB-I00 funded by the Spanish Ministry of Science, Innovation and Universities MCIU /AEI /10.13039/501100011033 and EU ERDF.

%%%%%%%%% -- BIB STYLE AND FILE -- %%%%%%%%
\bibliographystyle{IEEEtran}
\bibliography{refsv2}

@misc{champsim_paper,
	title        = {The Championship Simulator: Architectural Simulation for Education and Competition},
	author       = {Gober, Nathan and Chacon, Gino and Wang, Lei and Gratz, Paul V. and Jimenez, Daniel A. and Teran, Elvira and Pugsley, Seth and Kim, Jinchun},
	year         = 2022,
	publisher    = {arXiv},
	doi          = {10.48550/ARXIV.2210.14324},
	url          = {https://arxiv.org/abs/2210.14324},
	copyright    = {Creative Commons Attribution Share Alike 4.0 International}
}

@misc{champsim_cite,
	title        = {{ChampSim}},
	note         = {Accessed: 17-04-2024},
	howpublished = {\url{https://crc2.ece.tamu.edu/}}
}

@misc{arm_neoverse_v2_chipsncheese,
	title        = {{Hot Chips 2023: Arm’s Neoverse V2}},
	author       = {},
	howpublished = {\url{https://chipsandcheese.com/2023/09/11/hot-chips-2023-arms-neoverse-v2/}}
}

@misc{fnlmma,
	title        = {{The FNL+MMA Instruction Cache Prefetcher}},
	author       = {A. Seznec},
	howpublished = {\url{https://hal.inria.fr/hal-02884880/document}}
}

@misc{abishek_patterson_appendix,
	title        = {{Advanced Concepts on Address Translation, Appendix L in "Computer Architecture: A Quantitative Approach" by Hennessy and Patterson}},
	author       = {{Abhishek Bhattacharjee}},
	howpublished = {\url{http://www.cs.yale.edu/homes/abhishek/abhishek-appendix-l.pdf}}
}

@misc{cvp1,
	title        = {{Championship Value Prediction (CVP)}},
	note         = {Accessed: 17-04-2024},
	howpublished = {\url{https://www.microarch.org/cvp1/}}
}

@misc{ipc1,
	title        = {{The 1st Instruction Prefetching Championship}},
	note         = {Accessed: 17-04-2024},
	howpublished = {\url{https://research.ece.ncsu.edu/ipc/}}
}

@misc{a55,
	title        = {{ARM Cortex-A55 Core Technical Reference Manual r1p0}},
	howpublished = {\url{https://developer.arm.com/documentation/100442/0100/functional-description/level-1-memory-system/data-prefetching?lang=en}}
}

@article{Baer:1995,
	title        = {{Effective Hardware-Based Data Prefetching for High-Performance Processors}},
	author       = {Baer, Jean-Loup and Chen, Tien-Fu},
	year         = 1995,
	month        = {may},
	journal      = {IEEE Trans. Comput.},
	publisher    = {IEEE Computer Society},
	address      = {Washington, DC, USA},
	volume       = 44,
	number       = 5,
	pages        = {609--623},
	doi          = {10.1109/12.381947},
	issn         = {0018-9340},
	url          = {https://doi.org/10.1109/12.381947},
	issue_date   = {May 1995},
	numpages     = 15,
	acmid        = 627012
}

@article{llinux,
	title        = {{Using Hugetlbfs for Mapping Application Text Regions}},
	author       = {Doshi, Kshitij and Tran, Jantz},
	year         = 2006,
	month        = {01},
	pages        = {}
}

@article{asmdb,
	title        = {{AsmDB: Understanding and Mitigating Front-End Stalls in Warehouse-Scale Computers}},
	author       = {N. P. {Nagendra} and G. {Ayers} and D. I. {August} and H. K. {Cho} and S. {Kanev} and C. {Kozyrakis} and T. {Krishnamurthy} and H. {Litz} and T. {Moseley} and P. {Ranganathan}},
	year         = 2020,
	journal      = {IEEE Micro},
	volume       = 40,
	number       = 3,
	pages        = {56--63},
	doi          = {10.1145/3307650.3322234},
	url          = {https://doi.org/10.1145/3307650.3322234}
}

@inproceedings{Kandiraju:2002,
	title        = {{Going the Distance for TLB Prefetching: An Application-driven Study}},
	author       = {Kandiraju, Gokul B. and Sivasubramaniam, Anand},
	year         = 2002,
	booktitle    = {Proceedings of the 29th  International Symposium on Computer Architecture},
	location     = {Anchorage, Alaska},
	publisher    = {IEEE Computer Society},
	address      = {Washington, DC, USA},
	series       = {ISCA '02},
	pages        = {195--206},
	isbn         = {0-7695-1605-X},
	url          = {http://dl.acm.org/citation.cfm?id=545215.545237},
	numpages     = 12,
	acmid        = 545237
}

@inproceedings{Saulsbury:2000,
	title        = {{Recency-based TLB Preloading}},
	author       = {Saulsbury, Ashley and Dahlgren, Fredrik and Stenstr\"{o}m, Per},
	year         = 2000,
	booktitle    = {Proceedings of the 27th  International Symposium on Computer Architecture},
	location     = {Vancouver, British Columbia, Canada},
	publisher    = {ACM},
	address      = {New York, NY, USA},
	series       = {ISCA '00},
	pages        = {117--127},
	doi          = {10.1145/339647.339666},
	isbn         = {1-58113-232-8},
	url          = {http://doi.acm.org/10.1145/339647.339666},
	numpages     = 11,
	acmid        = 339666
}

@inproceedings{Pham:2015,
	title        = {{Large Pages and Lightweight Memory Management in Virtualized Environments: Can You Have It Both Ways?}},
	author       = {Pham, Binh and Vesel\'{y}, J\'{a}n and Loh, Gabriel H. and Bhattacharjee, Abhishek},
	year         = 2015,
	booktitle    = {Proceedings of the 48th International Symposium on Microarchitecture},
	location     = {Waikiki, Hawaii},
	publisher    = {ACM},
	address      = {New York, NY, USA},
	series       = {MICRO '15},
	pages        = {1--12},
	doi          = {10.1145/2830772.2830773},
	isbn         = {978-1-4503-4034-2},
	url          = {http://doi.acm.org/10.1145/2830772.2830773},
	numpages     = 12,
	acmid        = 2830773
}

@inproceedings{pom_tlb,
	title        = {{Rethinking TLB Designs in Virtualized Environments: A Very Large Part-of-Memory TLB}},
	author       = {Ryoo, Jee Ho and Gulur, Nagendra and Song, Shuang and John, Lizy K.},
	year         = 2017,
	booktitle    = {Proceedings of the 44th  International Symposium on Computer Architecture},
	location     = {Toronto, ON, Canada},
	publisher    = {ACM},
	address      = {New York, NY, USA},
	series       = {ISCA '17},
	pages        = {469--480},
	doi          = {10.1145/3079856.3080210},
	isbn         = {978-1-4503-4892-8},
	url          = {http://doi.acm.org/10.1145/3079856.3080210},
	numpages     = 12,
	acmid        = 3080210
}

@inproceedings{Alam:2017,
	title        = {{Do-It-Yourself Virtual Memory Translation}},
	author       = {Alam, Hanna and Zhang, Tianhao and Erez, Mattan and Etsion, Yoav},
	year         = 2017,
	booktitle    = {Proceedings of the 44th  International Symposium on Computer Architecture},
	location     = {Toronto, ON, Canada},
	publisher    = {ACM},
	address      = {New York, NY, USA},
	series       = {ISCA '17},
	pages        = {457--468},
	doi          = {10.1145/3079856.3080209},
	isbn         = {978-1-4503-4892-8},
	url          = {http://doi.acm.org/10.1145/3079856.3080209},
	numpages     = 12,
	acmid        = 3080209
}

@inproceedings{cloudsuite,
	title        = {{Clearing the Clouds: A Study of Emerging Scale-out Workloads on Modern Hardware}},
	author       = {Ferdman, Michael and Adileh, Almutaz and Kocberber, Onur and Volos, Stavros and Alisafaee, Mohammad and Jevdjic, Djordje and Kaynak, Cansu and Popescu, Adrian Daniel and Ailamaki, Anastasia and Falsafi, Babak},
	year         = 2012,
	booktitle    = {Proceedings of the 17th International Conference on Architectural Support for Programming Languages and Operating Systems},
	location     = {London, England, UK},
	publisher    = {ACM},
	address      = {New York, NY, USA},
	series       = {ASPLOS '12},
	pages        = {37--48},
	doi          = {10.1145/2150976.2150982},
	isbn         = {978-1-4503-0759-8},
	url          = {http://doi.acm.org/10.1145/2150976.2150982},
	numpages     = 12,
	acmid        = 2150982
}

@inproceedings{Kanev:2015:PWC,
	title        = {{Profiling a Warehouse-scale Computer}},
	author       = {Kanev, Svilen and Darago, Juan Pablo and Hazelwood, Kim and Ranganathan, Parthasarathy and Moseley, Tipp and Wei, Gu-Yeon and Brooks, David},
	year         = 2015,
	booktitle    = {Proceedings of the 42nd  International Symposium on Computer Architecture},
	location     = {Portland, Oregon},
	publisher    = {ACM},
	address      = {New York, NY, USA},
	series       = {ISCA '15},
	pages        = {158--169},
	doi          = {10.1145/2749469.2750392},
	isbn         = {978-1-4503-3402-0},
	url          = {http://doi.acm.org/10.1145/2749469.2750392},
	numpages     = 12,
	acmid        = 2750392
}

@inproceedings{gratz2020barca,
	title        = {BARCA: Branch-Agnostic Region Searching Algorithm},
	author       = {Paul Gratz and Daniel A. Jim{\'e}nez and Nathan Gober and Gino Chacon},
	year         = 2020,
	booktitle    = {Proceedings of the First Instruction Prefetching Championship (IPC)}
}

@inproceedings{ranger,
	title        = {{Translation Ranger: Operating System Support for Contiguity-Aware TLBs}},
	author       = {Yan, Zi and Lustig, Daniel and Nellans, David and Bhattacharjee, Abhishek},
	year         = 2019,
	booktitle    = {Proceedings of the 46th International Symposium on Computer Architecture},
	location     = {Phoenix, Arizona},
	publisher    = {Association for Computing Machinery},
	address      = {New York, NY, USA},
	series       = {ISCA '19},
	pages        = {698--710},
	doi          = {10.1145/3307650.3322223},
	isbn         = 9781450366694,
	url          = {https://doi.org/10.1145/3307650.3322223},
	numpages     = 13
}

@inproceedings{10.555,
	title        = {Coordinated and Efficient Huge Page Management with Ingens},
	author       = {Youngjin Kwon and Hangchen Yu and Simon Peter and Christopher J. Rossbach and Emmett Witchel},
	year         = 2016,
	month        = {nov},
	booktitle    = {12th USENIX Symposium on Operating Systems Design and Implementation (OSDI 16)},
	publisher    = {USENIX Association},
	address      = {Savannah, GA},
	pages        = {705--721},
	isbn         = {978-1-931971-33-1},
	url          = {https://www.usenix.org/conference/osdi16/technical-sessions/presentation/kwon}
}

@inproceedings{10.1145,
	title        = {Shared Address Translation Revisited},
	author       = {Dong, Xiaowan and Dwarkadas, Sandhya and Cox, Alan L.},
	year         = 2016,
	booktitle    = {Proceedings of the 11th European Conference on Computer Systems},
	location     = {London, United Kingdom},
	publisher    = {Association for Computing Machinery},
	address      = {New York, NY, USA},
	series       = {EuroSys '16},
	doi          = {10.1145/2901318.2901327},
	isbn         = 9781450342407,
	url          = {https://doi.org/10.1145/2901318.2901327},
	articleno    = {Article 18},
	numpages     = 15
}

@inproceedings{bolt,
	title        = {{BOLT: A Practical Binary Optimizer for Data Centers and Beyond}},
	author       = {Panchenko, Maksim and Auler, Rafael and Nell, Bill and Ottoni, Guilherme},
	year         = 2019,
	booktitle    = {Proceedings of the 2019 International Symposium on Code Generation and Optimization},
	location     = {Washington, DC, USA},
	publisher    = {IEEE Press},
	series       = {CGO '19},
	pages        = {2--14},
	doi          = {10.1109/CGO.2019.8661201},
	isbn         = 9781728114361,
	url          = {https://doi.org/10.1109/CGO.2019.8661201},
	numpages     = 13
}

@inproceedings{epi_isca,
	title        = {{A Cost-Effective Entangling Prefetcher for Instructions}},
	author       = {Ros, Alberto and Jimborean, Alexandra},
	year         = 2021,
	booktitle    = {Proceedings of the 48th  International Symposium on Computer Architecture},
	series       = {ISCA '21},
	volume       = {},
	number       = {},
	pages        = {99--111},
	doi          = {10.1109/ISCA52012.2021.00017},
	url          = {https://doi.org/10.1109/ISCA52012.2021.00017}
}

@inproceedings{chirp,
	title        = {{CHiRP: Control-Flow History Reuse Prediction}},
	author       = {S. {Mirbagher-Ajorpaz} and E. {Garza} and G. {Pokam} and D. A. {Jim\'{e}nez}},
	year         = 2020,
	booktitle    = {Proceedings of the 2020 53rd  International Symposium on Microarchitecture},
	series       = {MICRO '16},
	volume       = {},
	number       = {},
	pages        = {131--145},
	doi          = {10.1109/MICRO50266.2020.00023},
	url          = {https://doi.org/10.1109/MICRO50266.2020.00023}
}

@inproceedings{diligent_tlbs,
	title        = {{Diligent TLBs: A Mechanism for Exploiting Heterogeneity in TLB Miss Behavior}},
	author       = {Elnawawy, Hussein and Chowdhury, Rangeen Basu Roy and Awad, Amro and Byrd, Gregory T.},
	year         = 2019,
	booktitle    = {Proceedings of the International Conference on Supercomputing},
	location     = {Phoenix, Arizona},
	publisher    = {Association for Computing Machinery},
	address      = {New York, NY, USA},
	series       = {ICS '19},
	pages        = {195–205},
	doi          = {10.1145/3330345.3330363},
	isbn         = 9781450360791,
	url          = {https://doi-org.recursos.biblioteca.upc.edu/10.1145/3330345.3330363},
	numpages     = 11
}

@inproceedings{multiperspective,
	title        = {{Multiperspective Reuse Prediction}},
	author       = {Jim\'{e}nez, Daniel A. and Teran, Elvira},
	year         = 2017,
	booktitle    = {Proceedings of the 50th  International Symposium on Microarchitecture},
	location     = {Cambridge, Massachusetts},
	publisher    = {Association for Computing Machinery},
	address      = {New York, NY, USA},
	series       = {MICRO '17},
	pages        = {436–448},
	doi          = {10.1145/3123939.3123942},
	isbn         = 9781450349529,
	url          = {https://doi.org/10.1145/3123939.3123942},
	numpages     = 13
}

@inproceedings{osdi21,
	title        = {{Beyond malloc Efficiency to Fleet Efficiency: a Hugepage-aware Memory Allocator}},
	author       = {A.H. Hunter and Chris Kennelly and Paul Turner and Darryl Gove and Tipp Moseley and Parthasarathy Ranganathan},
	year         = 2021,
	month        = {jul},
	booktitle    = {Proceedings of the 15th {USENIX} Symposium on Operating Systems Design and Implementation},
	publisher    = {{USENIX} Association},
	series       = {OSDI '21},
	pages        = {257--273},
	isbn         = {978-1-939133-22-9},
	url          = {https://www.usenix.org/conference/osdi21/presentation/hunter}
}

@inproceedings{6237026,
	title        = {{Reducing Memory Reference Energy with Opportunistic Virtual Caching}},
	author       = {Basu, Arkaprava and Hill, Mark D. and Swift, Michael M.},
	year         = 2012,
	booktitle    = {Proceedings of the 39th  International Symposium on Computer Architecture},
	series       = {ISCA '12},
	volume       = {},
	number       = {},
	pages        = {297--308},
	doi          = {10.1109/ISCA.2012.6237026},
	url          = {https://doi.org/10.1109/ISCA.2012.6237026}
}

@book{10.5555/77493,
	title        = {{Computer Architecture: A Quantitative Approach}},
	author       = {Patterson, David A. and Hennessy, John L.},
	year         = 1990,
	publisher    = {Morgan Kaufmann Publishers Inc.},
	address      = {San Francisco, CA, USA},
	isbn         = 1558800698
}

@inproceedings{berti,
	title        = {Berti: an Accurate Local-Delta Data Prefetcher},
	author       = {Navarro-Torres, Agustín and Panda, Biswabandan and Alastruey-Benedé, Jesús and Ibáñez, Pablo and Viñals-Yúfera, Víctor and Ros, Alberto},
	year         = 2022,
	booktitle    = {Proceedings of the 55th International Symposium on Microarchitecture},
	series       = {MICRO '22},
	volume       = {},
	number       = {},
	pages        = {975--991},
	doi          = {10.1109/MICRO56248.2022.00072},
	url          = {https://doi.org/10.1109/MICRO56248.2022.00072}
}

@inproceedings{atp_sbfp,
	title        = {Exploiting Page Table Locality for Agile TLB Prefetching},
	author       = {Vavouliotis, Georgios and Alvarez, Lluc and Karakostas, Vasileios and Nikas, Konstantinos and Koziris, Nectarios and Jiménez, Daniel A. and Casas, Marc},
	year         = 2021,
	booktitle    = {2021 ACM/IEEE 48th Annual International Symposium on Computer Architecture (ISCA)},
	volume       = {},
	number       = {},
	pages        = {85--98},
	doi          = {10.1109/ISCA52012.2021.00016},
	url          = {https://doi.org/10.1109/ISCA52012.2021.00016}
}

@inproceedings{morrigan,
	title        = {{Morrigan: A Composite Instruction TLB Prefetcher}},
	author       = {Vavouliotis, Georgios and Alvarez, Lluc and Grot, Boris and Jim\'{e}nez, Daniel and Casas, Marc},
	year         = 2021,
	booktitle    = {Proceedings of the 54th  International Symposium on Microarchitecture},
	location     = {Virtual Event, Greece},
	publisher    = {Association for Computing Machinery},
	address      = {New York, NY, USA},
	series       = {MICRO '21},
	pages        = {1138–1153},
	doi          = {10.1145/3466752.3480049},
	isbn         = 9781450385572,
	url          = {https://doi.org/10.1145/3466752.3480049},
	numpages     = 16
}

@inproceedings{psa,
	title        = {{Page Size Aware Cache Prefetching}},
	author       = {Vavouliotis, Georgios and Chacon, Gino and Alvarez, Lluc and Gratz, Paul V. and Jiménez, Daniel A. and Casas, Marc},
	year         = 2022,
	booktitle    = {Proceedings of the 55th International Symposium on Microarchitecture},
	series       = {MICRO '22},
	volume       = {},
	number       = {},
	pages        = {956--974},
	doi          = {10.1109/MICRO56248.2022.00070},
	url          = {https://doi.org/10.1109/MICRO56248.2022.00070}
}

@inproceedings{decoupled_frontend_arm,
	title        = {Re-establishing Fetch-Directed Instruction Prefetching: An Industry Perspective},
	author       = {Ishii, Yasuo and Lee, Jaekyu and Nathella, Krishnendra and Sunwoo, Dam},
	year         = 2021,
	booktitle    = {2021 IEEE International Symposium on Performance Analysis of Systems and Software (ISPASS)},
	volume       = {},
	number       = {},
	pages        = {172--182},
	doi          = {10.1109/ISPASS51385.2021.00034},
	url          = {https://doi.org/10.1109/ISPASS51385.2021.00034}
}

@inproceedings{fdip,
	title        = {{Fetch Directed Instruction Prefetching}},
	author       = {Reinman, G. and Calder, B. and Austin, T.},
	year         = 1999,
	booktitle    = {Proceedings of the 32nd   International Symposium on Microarchitecture},
	series       = {MICRO '99},
	volume       = {},
	number       = {},
	pages        = {16--27},
	doi          = {10.1109/MICRO.1999.809439},
	url          = {https://doi.org/10.1109/MICRO.1999.809439}
}

@article{skylake_uarch,
	title        = {Inside 6th-Generation Intel Core: New Microarchitecture Code-Named Skylake},
	author       = {Doweck, Jack and Kao, Wen-Fu and Lu, Allen Kuan-yu and Mandelblat, Julius and Rahatekar, Anirudha and Rappoport, Lihu and Rotem, Efraim and Yasin, Ahmad and Yoaz, Adi},
	year         = 2017,
	journal      = {IEEE Micro},
	volume       = 37,
	number       = 2,
	pages        = {52--62},
	doi          = {10.1109/MM.2017.38},
	url          = {https://doi.org/10.1109/MM.2017.38}
}

@inproceedings{victima,
	title        = {Victima: Drastically Increasing Address Translation Reach by Leveraging Underutilized Cache Resources},
	author       = {Kanellopoulos, Konstantinos and Nam, Hong Chul and Bostanci, Nisa and Bera, Rahul and Sadrosadati, Mohammad and Kumar, Rakesh and Bartolini, Davide Basilio and Mutlu, Onur},
	year         = 2023,
	booktitle    = {Proceedings of the 56th Annual IEEE/ACM International Symposium on Microarchitecture},
	location     = {, Toronto, ON, Canada,},
	publisher    = {Association for Computing Machinery},
	address      = {New York, NY, USA},
	series       = {MICRO '23},
	pages        = {1178–1195},
	doi          = {10.1145/3613424.3614276},
	isbn         = 9798400703294,
	url          = {https://doi.org/10.1145/3613424.3614276},
	numpages     = 18
}

@inproceedings{teran2016perceptron,
	title        = {Perceptron learning for reuse prediction},
	author       = {Teran, Elvira and Wang, Zhe and Jim{\'e}nez, Daniel A},
	year         = 2016,
	booktitle    = {2016 49th Annual IEEE/ACM International Symposium on Microarchitecture (MICRO)},
	pages        = {1--12},
	doi          = {10.1109/MICRO.2016.7783705},
	url          = {https://doi.org/10.1109/MICRO.2016.7783705},
	organization = {IEEE}
}

@inproceedings{pacman,
	title        = {PACMan: Prefetch-Aware Cache Management for high performance caching},
	author       = {Wu, Carole-Jean and Jaleel, Aamer and Martonosi, Margaret and Steely, Simon C. and Emer, Joel},
	year         = 2011,
	booktitle    = {2011 44th Annual IEEE/ACM International Symposium on Microarchitecture (MICRO)},
	volume       = {},
	number       = {},
	pages        = {442--453},
	doi          = {10.1145/2155620.215567},
	url          = {https://doi.org/10.1145/2155620.215567}
}

@inproceedings{hawkeye,
	title        = {Back to the Future: Leveraging Belady's Algorithm for Improved Cache Replacement},
	author       = {Jain, Akanksha and Lin, Calvin},
	year         = 2016,
	booktitle    = {2016 ACM/IEEE 43rd Annual International Symposium on Computer Architecture (ISCA)},
	volume       = {},
	number       = {},
	pages        = {78--89},
	doi          = {10.1109/ISCA.2016.17},
	url          = {https://doi.org/10.1109/ISCA.2016.17}
}

@inproceedings{bera_hermes,
	title        = {Hermes: Accelerating Long-Latency Load Requests via Perceptron-Based Off-Chip Load Prediction},
	author       = {Bera, Rahul and Kanellopoulos, Konstantinos and Balachandran, Shankar and Novo, David and Olgun, Ataberk and Sadrosadat, Mohammad and Mutlu, Onur},
	year         = 2022,
	month        = {Oct},
	booktitle    = {2022 55th IEEE/ACM International Symposium on Microarchitecture (MICRO)},
	volume       = {},
	number       = {},
	pages        = {1--18},
	doi          = {10.1109/MICRO56248.2022.00015},
	issn         = {},
	url          = {https://doi.org/10.1109/MICRO56248.2022.00015}
}

@inproceedings{jimenez_multiperspective_2017,
	title        = {Multiperspective reuse prediction},
	author       = {Jim{\'e}nez, Daniel A and Teran, Elvira},
	year         = 2017,
	booktitle    = {2017 50th Annual IEEE/ACM International Symposium on Microarchitecture (MICRO)},
	location     = {Cambridge, Massachusetts},
	publisher    = {Association for Computing Machinery},
	address      = {New York, NY, USA},
	series       = {MICRO-50 '17},
	pages        = {436--448},
	doi          = {10.1145/3123939.3123942},
	isbn         = 9781450349529,
	url          = {https://doi.org/10.1145/3123939.3123942},
	organization = {IEEE},
	numpages     = 13
}

@inproceedings{10.1145/3669940.3707247,
	title        = {Instruction-Aware Cooperative TLB and Cache Replacement Policies},
	author       = {Chasapis, Dimitrios and Vavouliotis, Georgios and Jim\'{e}nez, Daniel A. and Casas, Marc},
	year         = 2025,
	booktitle    = {Proceedings of the 30th ACM International Conference on Architectural Support for Programming Languages and Operating Systems, Volume 1},
	location     = {Rotterdam, Netherlands},
	publisher    = {Association for Computing Machinery},
	address      = {New York, NY, USA},
	series       = {ASPLOS '25},
	pages        = {619–636},
	doi          = {10.1145/3669940.3707247},
	isbn         = 9798400706981,
	url          = {https://doi.org/10.1145/3669940.3707247},
	numpages     = 18
}

@inproceedings{schall2024llbp,
	title        = {The Last-Level Branch Predictor},
	author       = {Schall, David and Sandberg, Andreas and Grot, Boris},
	year         = 2024,
	booktitle    = {2024 57th IEEE/ACM International Symposium on Microarchitecture (MICRO)},
	volume       = {},
	number       = {},
	pages        = {464--479},
	doi          = {10.1109/MICRO61859.2024.00042},
	url          = {https://doi.org/10.1109/MICRO61859.2024.00042}
}

@inproceedings{9773195,
	title        = {Effective Mimicry of Belady’s MIN Policy},
	author       = {Shah, Ishan and Jain, Akanksha and Lin, Calvin},
	year         = 2022,
	month        = {April},
	booktitle    = {2022 IEEE International Symposium on High-Performance Computer Architecture (HPCA)},
	volume       = {},
	number       = {},
	pages        = {558--572},
	doi          = {10.1109/HPCA53966.2022.00048},
	issn         = {2378-203X},
	url          = {https://doi.org/10.1109/HPCA53966.2022.00048}
}

@inproceedings{ship,
	title        = {SHiP: signature-based hit predictor for high performance caching},
	author       = {Wu, Carole-Jean and Jaleel, Aamer and Hasenplaugh, Will and Martonosi, Margaret and Steely, Simon C. and Emer, Joel},
	year         = 2011,
	booktitle    = {Proceedings of the 44th Annual IEEE/ACM International Symposium on Microarchitecture},
	location     = {Porto Alegre, Brazil},
	publisher    = {Association for Computing Machinery},
	address      = {New York, NY, USA},
	series       = {MICRO-44},
	pages        = {430–441},
	doi          = {10.1145/2155620.2155671},
	isbn         = 9781450310536,
	url          = {https://doi.org/10.1145/2155620.2155671},
	numpages     = 12
}

@inproceedings{5470352,
	title        = {Adapting cache partitioning algorithms to pseudo-LRU replacement policies},
	author       = {Kedzierski, Kamil and Moreto, Miquel and Cazorla, Francisco J. and Valero, Mateo},
	year         = 2010,
	booktitle    = {2010 IEEE International Symposium on Parallel \& Distributed Processing (IPDPS)},
	volume       = {},
	number       = {},
	pages        = {1--12},
	doi          = {10.1109/IPDPS.2010.5470352},
	url          = {https://doi.org/10.1109/IPDPS.2010.5470352}
}

@inproceedings{10.1145/2540708.2540733,
	title        = {Insertion and promotion for tree-based PseudoLRU last-level caches},
	author       = {Jim\'{e}nez, Daniel A.},
	year         = 2013,
	booktitle    = {Proceedings of the 46th Annual IEEE/ACM International Symposium on Microarchitecture},
	location     = {Davis, California},
	publisher    = {Association for Computing Machinery},
	address      = {New York, NY, USA},
	series       = {MICRO-46},
	pages        = {284–296},
	doi          = {10.1145/2540708.2540733},
	isbn         = 9781450326384,
	url          = {https://doi.org/10.1145/2540708.2540733},
	numpages     = 13
}

@patent{chen_implementation_2006,
	title        = {Implementation of a pseudo-{LRU} algorithm in a partitioned cache},
	author       = {Chen, Wen-Tzer Thomas and Liu, Peichun Peter and Stelzer, Kevin C.},
	year         = 2006,
	month        = {jun},
	number       = {US7069390B2},
	url          = {https://patents.google.com/patent/US7069390B2/en},
	urldate      = {2025-04-01},
	nationality  = {US},
	language     = {en},
	assignee     = {International Business Machines Corp}
}

@inproceedings{thermometer,
	title        = {Thermometer: profile-guided btb replacement for data center applications},
	author       = {Song, Shixin and Khan, Tanvir Ahmed and Shahri, Sara Mahdizadeh and Sriraman, Akshitha and Soundararajan, Niranjan K and Subramoney, Sreenivas and Jim\'{e}nez, Daniel A. and Litz, Heiner and Kasikci, Baris},
	year         = 2022,
	booktitle    = {Proceedings of the 49th Annual International Symposium on Computer Architecture},
	location     = {New York, New York},
	publisher    = {Association for Computing Machinery},
	address      = {New York, NY, USA},
	series       = {ISCA '22},
	pages        = {742–756},
	doi          = {10.1145/3470496.3527430},
	isbn         = 9781450386104,
	url          = {https://doi.org/10.1145/3470496.3527430},
	numpages     = 15
}

@inproceedings{Gao2010ADS,
  author    = {Hongliang Gao and Chris Wilkerson},
  title     = {{A Dueling Segmented LRU Replacement Algorithm with Adaptive Bypassing}},
  booktitle = {1st JILP Workshop on Computer Architecture Competitions (JWAC-1): Cache Replacement Championship},
  year      = {2010},
  month     = jun,
  location  = {Saint-Malo, France},
  url       = {https://jilp.org/jwac-1/online/papers/005_gao.pdf}
}

@inproceedings{srrip,
	title        = {High Performance Cache Replacement Using Re-Reference Interval Prediction (RRIP)},
	author       = {Jaleel, Aamer and Theobald, Kevin B. and Steely, Simon C. and Emer, Joel},
	year         = 2010,
	booktitle    = {Proceedings of the 37th Annual International Symposium on Computer Architecture},
	location     = {Saint-Malo, France},
	publisher    = {Association for Computing Machinery},
	address      = {New York, NY, USA},
	series       = {ISCA '10},
	pages        = {60–71},
	doi          = {10.1145/1815961.1815971},
	isbn         = 9781450300537,
	url          = {https://doi.org/10.1145/1815961.1815971},
	numpages     = 12
}

@inproceedings{10.1145/301453.301487,
	title        = {On the Existence of a Spectrum of Policies That Subsumes the Least Recently Used (LRU) and Least Frequently Used (LFU) Policies},
	author       = {Lee, Donghee and Choi, Jongmoo and Kim, Jong-Hun and Noh, Sam H. and Min, Sang Lyul and Cho, Yookun and Kim, Chong Sang},
	year         = 1999,
	booktitle    = {Proceedings of the 1999 ACM SIGMETRICS International Conference on Measurement and Modeling of Computer Systems},
	location     = {Atlanta, Georgia, USA},
	publisher    = {Association for Computing Machinery},
	address      = {New York, NY, USA},
	series       = {SIGMETRICS '99},
	pages        = {134–143},
	doi          = {10.1145/301453.301487},
	isbn         = {158113083X},
	url          = {https://doi.org/10.1145/301453.301487},
	numpages     = 10
}

@inproceedings{10.1145/170035.170081,
	title        = {The LRU-K Page Replacement Algorithm for Database Disk Buffering},
	author       = {O'Neil, Elizabeth J. and O'Neil, Patrick E. and Weikum, Gerhard},
	year         = 1993,
	booktitle    = {Proceedings of the 1993 ACM SIGMOD International Conference on Management of Data},
	location     = {Washington, D.C., USA},
	publisher    = {Association for Computing Machinery},
	address      = {New York, NY, USA},
	series       = {SIGMOD '93},
	pages        = {297–306},
	doi          = {10.1145/170035.170081},
	isbn         = {0897915925},
	url          = {https://doi.org/10.1145/170035.170081},
	numpages     = 10
}

@inproceedings{10.1145/1250662.1250709,
	title        = {Adaptive Insertion Policies for High Performance Caching},
	author       = {Qureshi, Moinuddin K. and Jaleel, Aamer and Patt, Yale N. and Steely, Simon C. and Emer, Joel},
	year         = 2007,
	booktitle    = {Proceedings of the 34th Annual International Symposium on Computer Architecture},
	location     = {San Diego, California, USA},
	publisher    = {Association for Computing Machinery},
	address      = {New York, NY, USA},
	series       = {ISCA '07},
	pages        = {381–391},
	doi          = {10.1145/1250662.1250709},
	isbn         = 9781595937063,
	url          = {https://doi.org/10.1145/1250662.1250709},
	numpages     = 11
}

@inproceedings{4041862,
	title        = {Adaptive Caches: Effective Shaping of Cache Behavior to Workloads},
	author       = {Subramanian, Ranjith and Smaragdakis, Yannis and Loh, Gabriel H.},
	year         = 2006,
	month        = {Dec},
	booktitle    = {2006 39th Annual IEEE/ACM International Symposium on Microarchitecture (MICRO'06)},
	volume       = {},
	number       = {},
	pages        = {385--396},
	doi          = {10.1109/MICRO.2006.7},
	issn         = {2379-3155},
	url          = {https://doi.org/10.1109/MICRO.2006.7}
}

@inproceedings{824338,
	title        = {Modified LRU policies for improving second-level cache behavior},
	author       = {Wong, W.A. and Baer, J.-L.},
	year         = 2000,
	booktitle    = {Proceedings Sixth International Symposium on High-Performance Computer Architecture. HPCA-6 (Cat. No.PR00550)},
	volume       = {},
	number       = {},
	pages        = {49--60},
	doi          = {10.1109/HPCA.2000.824338},
	url          = {https://doi.org/10.1109/HPCA.2000.824338}
}

@inproceedings{glider,
	title        = {Applying Deep Learning to the Cache Replacement Problem},
	author       = {Shi, Zhan and Huang, Xiangru and Jain, Akanksha and Lin, Calvin},
	year         = 2019,
	booktitle    = {Proceedings of the 52nd Annual IEEE/ACM International Symposium on Microarchitecture},
	location     = {Columbus, OH, USA},
	publisher    = {Association for Computing Machinery},
	address      = {New York, NY, USA},
	series       = {MICRO '52},
	pages        = {413–425},
	doi          = {10.1145/3352460.3358319},
	isbn         = 9781450369381,
	url          = {https://doi.org/10.1145/3352460.3358319},
	numpages     = 13
}

@inproceedings{5695535,
	title        = {Sampling Dead Block Prediction for Last-Level Caches},
	author       = {Khan, Samira Manabi and Tian, Yingying and Jiménez, Daniel A.},
	year         = 2010,
	month        = {Dec},
	booktitle    = {2010 43rd Annual IEEE/ACM International Symposium on Microarchitecture},
	volume       = {},
	number       = {},
	pages        = {175--186},
	doi          = {10.1109/MICRO.2010.24},
	issn         = {2379-3155},
	url          = {https://doi.org/10.1109/MICRO.2010.24}
}

@inproceedings{10.1109/MICRO.2012.43,
	title        = {Improving Cache Management Policies Using Dynamic Reuse Distances},
	author       = {Duong, Nam and Zhao, Dali and Kim, Taesu and Cammarota, Rosario and Valero, Mateo and Veidenbaum, Alexander V.},
	year         = 2012,
	booktitle    = {Proceedings of the 2012 45th Annual IEEE/ACM International Symposium on Microarchitecture},
	location     = {Vancouver, B.C., CANADA},
	publisher    = {IEEE Computer Society},
	address      = {USA},
	series       = {MICRO-45},
	pages        = {389–400},
	doi          = {10.1109/MICRO.2012.43},
	isbn         = 9780769549248,
	url          = {https://doi.org/10.1109/MICRO.2012.43},
	numpages     = 12
}

@inproceedings{10.5555/545215.545239,
	title        = {Timekeeping in the Memory System: Predicting and Optimizing Memory Behavior},
	author       = {Hu, Zhigang and Kaxiras, Stefanos and Martonosi, Margaret},
	year         = 2002,
	booktitle    = {Proceedings of the 29th Annual International Symposium on Computer Architecture},
	location     = {Anchorage, Alaska},
	publisher    = {IEEE Computer Society},
	address      = {USA},
	series       = {ISCA '02},
	pages        = {209–220},
	doi          = {10.1145/545214.545239},
	isbn         = {076951605X},
	url          = {https://doi.org/10.1145/545214.545239},
	numpages     = 12
}

@inproceedings{4601909,
	title        = {Cache replacement based on reuse-distance prediction},
	author       = {Keramidas, Georgios and Petoumenos, Pavlos and Kaxiras, Stefanos},
	year         = 2007,
	booktitle    = {2007 25th International Conference on Computer Design},
	volume       = {},
	number       = {},
	pages        = {245--250},
	doi          = {10.1109/ICCD.2007.4601909},
	url          = {https://doi.org/10.1109/ICCD.2007.4601909}
}

@article{4358260,
	title        = {Counter-Based Cache Replacement and Bypassing Algorithms},
	author       = {Kharbutli, Mazen and Solihin, Yan},
	year         = 2008,
	journal      = {IEEE Transactions on Computers},
	volume       = 57,
	number       = 4,
	pages        = {433--447},
	doi          = {10.1109/TC.2007.70816},
	url          = {https://doi.org/10.1109/TC.2007.70816}
}

@inproceedings{4771793,
	title        = {Cache bursts: A new approach for eliminating dead blocks and increasing cache efficiency},
	author       = {Liu, Haiming and Ferdman, Michael and Huh, Jaehyuk and Burger, Doug},
	year         = 2008,
	month        = {Nov},
	booktitle    = {2008 41st IEEE/ACM International Symposium on Microarchitecture},
	volume       = {},
	number       = {},
	pages        = {222--233},
	doi          = {10.1109/MICRO.2008.4771793},
	issn         = {2379-3155},
	url          = {https://doi.org/10.1109/MICRO.2008.4771793}
}

@inproceedings{8091244,
	title        = {Leeway: Addressing Variability in Dead-Block Prediction for Last-Level Caches},
	author       = {Faldu, Priyank and Grot, Boris},
	year         = 2017,
	booktitle    = {2017 26th International Conference on Parallel Architectures and Compilation Techniques (PACT)},
	volume       = {},
	number       = {},
	pages        = {180--193},
	doi          = {10.1109/PACT.2017.32},
	url          = {https://doi.org/10.1109/PACT.2017.32}
}

@inproceedings{emissary,
	title        = {EMISSARY: Enhanced Miss Awareness Replacement Policy for L2 Instruction Caching},
	author       = {Nagendra, Nayana Prasad and Godala, Bhargav Reddy and Chaturvedi, Ishita and Patel, Atmn and Kanev, Svilen and Moseley, Tipp and Stark, Jared and Pokam, Gilles A. and Campanoni, Simone and August, David I.},
	year         = 2023,
	booktitle    = {Proceedings of the 50th Annual International Symposium on Computer Architecture},
	location     = {Orlando, FL, USA},
	publisher    = {Association for Computing Machinery},
	address      = {New York, NY, USA},
	series       = {ISCA '23},
	doi          = {10.1145/3579371.3589097},
	isbn         = 9798400700958,
	url          = {https://doi.org/10.1145/3579371.3589097},
	articleno    = 62,
	numpages     = 13
}

@inproceedings{clip,
	title        = {High performing cache hierarchies for server workloads: Relaxing inclusion to capture the latency benefits of exclusive caches},
	author       = {Jaleel, Aamer and Nuzman, Joseph and Moga, Adrian and Steely, Simon C. and Emer, Joel},
	year         = 2015,
	booktitle    = {2015 IEEE 21st International Symposium on High Performance Computer Architecture (HPCA)},
	volume       = {},
	number       = {},
	pages        = {343--353},
	doi          = {10.1109/HPCA.2015.7056045},
	url          = {https://doi.org/10.1109/HPCA.2015.7056045}
}

@inproceedings{Young2017SHiP,
  author    = {Vinson Young and Chia-Chen Chou and Aamer Jaleel and Moinuddin K. Qureshi},
  title     = {{SHiP++: Enhancing Signature-Based Hit Predictor for Improved Cache Performance}},
  booktitle = {2nd Cache Replacement Championship (CRC-2), in conjunction with ISCA 2017},
  year      = {2017},
  month     = jun,
  location  = {Toronto, Canada},
  url       = {https://crc2.ece.tamu.edu/?page_id=53}
}

@inproceedings{paciv,
	title        = {Light-weight Cache Replacement for Instruction Heavy Workloads},
	author       = {Mostofi, Saba and Gupta, Setu and Hassani, Ahmad and Tibrewala, Krishnam and Teran, Elvira and Gratz, Paul V. and Jim\'{e}nez, Daniel A.},
	year         = 2025,
	booktitle    = {Proceedings of the 52nd Annual International Symposium on Computer Architecture},
	location     = {},
	publisher    = {Association for Computing Machinery},
	address      = {New York, NY, USA},
	series       = {ISCA '25},
	pages        = {1005–1019},
	doi          = {10.1145/3695053.3730993},
	isbn         = 9798400712616,
	url          = {https://doi.org/10.1145/3695053.3730993},
	numpages     = 15
}

@inproceedings{vavou25,
	title        = {To Cross, or Not to Cross Pages for Prefetching?},
	author       = {Vavouliotis, Georgios and Torrents, Marti and Grot, Boris and Kalaitzidis, Kleovoulos and Peled, Leeor and Casas, Marc},
	year         = 2025,
	month        = {March},
	booktitle    = {2025 IEEE International Symposium on High Performance Computer Architecture (HPCA)},
	volume       = {},
	number       = {},
	pages        = {188--203},
	doi          = {10.1109/HPCA61900.2025.00025},
	issn         = {2378-203X},
	url          = {https://doi.org/10.1109/HPCA61900.2025.00025}
}

@patent{amd_prefetch_flag_2023,
  title        = {Processing Metadata, Policies, and Composite Tags (Prefetch Flag in Cache Tag)},
  author       = {{Advanced Micro Devices, Inc.}},
  nationality  = {United States},
  year         = 2023,
  number       = {US11635960B2},
  url          = {https://patents.google.com/patent/US11635960B2},
  note         = {Describes a metadata flag in cache tags used to track or prevent prefetching into caches},
}

@manual{arm_pl310_trm,
	title        = {{PL310 L2 Cache Controller Technical Reference Manual}},
	year         = 2010,
	note         = {Section 3.3.6, "Prefetch Control Register." Available: \url{https://developer.arm.com/documentation/ddi0246/c/}},
	organization = {Arm Ltd.}
}

@manual{intel_sdm,
	title        = {{Intel$^\circledR$ 64 and IA-32 Architectures Software Developer’s Manual, Volume 3 (System Programming Guide)}},
	year         = 2023,
	note         = {Section 19.10, "Performance Monitoring Events for Intel\textregistered~Core\texttrademark~Processors." Available: \url{https://software.intel.com/content/www/us/en/develop/articles/intel-sdm.html}},
	organization = {Intel Corporation}
}

@inproceedings{7095786,
	title        = {Multi-program benchmark definition},
	author       = {Jacobvitz, Adam N. and Hilton, Andrew D. and Sorin, Daniel J.},
	year         = 2015,
	month        = {March},
	booktitle    = {2015 IEEE International Symposium on Performance Analysis of Systems and Software (ISPASS)},
	volume       = {},
	number       = {},
	pages        = {72--82},
	doi          = {10.1109/ISPASS.2015.7095786},
	issn         = {},
	url          = {https://doi.org/10.1109/ISPASS.2015.7095786}
}

@article{SeznecM06,
	title        = {A Case for (Partially) TAgged GEometric History Length Branch Prediction},
	author       = {Andr{\'e} Seznec and Pierre Michaud},
	year         = 2006,
	journal      = {Journal of Instruction‐Level Parallelism},
	volume       = 8,
	url          = {http://www.jilp.org/vol8/v8paper1.pdf},
	note         = {Special issue on Branch Prediction}
}

@inproceedings{tage,
	title        = {A new case for the TAGE branch predictor},
	author       = {Seznec, Andre},
	year         = 2011,
	booktitle    = {Proceedings of the 44th Annual IEEE/ACM International Symposium on Microarchitecture},
	location     = {Porto Alegre, Brazil},
	publisher    = {Association for Computing Machinery},
	address      = {New York, NY, USA},
	series       = {MICRO-44},
	pages        = {117–127},
	doi          = {10.1145/2155620.2155635},
	isbn         = 9781450310536,
	url          = {https://doi.org/10.1145/2155620.2155635},
	numpages     = 11
}

@inproceedings{10.5555/3049832.3049858,
	title        = {Optimizing function placement for large-scale data-center applications},
	author       = {Ottoni, Guilherme and Maher, Bertrand},
	year         = 2017,
	booktitle    = {Proceedings of the 2017 International Symposium on Code Generation and Optimization},
	location     = {Austin, USA},
	publisher    = {IEEE Press},
	series       = {CGO '17},
	pages        = {233–244},
	isbn         = 9781509049318,
	numpages     = 12
}

@inproceedings{codestitcher,
	title        = {Codestitcher: inter-procedural basic block layout optimization},
	author       = {Lavaee, Rahman and Criswell, John and Ding, Chen},
	year         = 2019,
	booktitle    = {Proceedings of the 28th International Conference on Compiler Construction},
	location     = {Washington, DC, USA},
	publisher    = {Association for Computing Machinery},
	address      = {New York, NY, USA},
	series       = {CC 2019},
	pages        = {65–75},
	doi          = {10.1145/3302516.3307358},
	isbn         = 9781450362771,
	url          = {https://doi.org/10.1145/3302516.3307358},
	numpages     = 11
}

@inproceedings{10476485,
	title        = {A Two Level Neural Approach Combining Off-Chip Prediction with Adaptive Prefetch Filtering},
	author       = {Jamet, Alexandre Valentin and Vavouliotis, Georgios and Jiménez, Daniel A. and Alvarez, Lluc and Casas, Marc},
	year         = 2024,
	booktitle    = {2024 IEEE International Symposium on High-Performance Computer Architecture (HPCA)},
	volume       = {},
	number       = {},
	pages        = {528--542},
	doi          = {10.1109/HPCA57654.2024.00046},
	url          = {https://doi.org/10.1109/HPCA57654.2024.00046}
}

@online{noauthor_cascade_nodate,
	title        = {Cascade Lake - Microarchitectures - Intel - {WikiChip}},
	url          = {https://en.wikichip.org/wiki/intel/microarchitectures/cascade_lake#Memory_Hierarchy},
	urldate      = {2021-06-29}
}

@inproceedings{seznec:hal-00639041,
	title        = {{A 64-Kbytes ITTAGE indirect branch predictor}},
	author       = {Seznec, Andr{\'e}},
	year         = 2011,
	month        = {Jun},
	booktitle    = {{JWAC-2: Championship Branch Prediction}},
	address      = {San Jose, United States},
	url          = {https://inria.hal.science/hal-00639041},
	organization = {{JILP}},
	hal_id       = {hal-00639041},
	hal_version  = {v1}
}

@inproceedings{makis1,
	title        = {Micro-Armed Bandit: Lightweight \& Reusable Reinforcement Learning for Microarchitecture Decision-Making},
	author       = {Gerogiannis, Gerasimos and Torrellas, Josep},
	year         = 2023,
	booktitle    = {Proceedings of the 56th Annual IEEE/ACM International Symposium on Microarchitecture},
	location     = {Toronto, ON, Canada},
	publisher    = {Association for Computing Machinery},
	address      = {New York, NY, USA},
	series       = {MICRO '23},
	pages        = {698–713},
	doi          = {10.1145/3613424.3623780},
	isbn         = 9798400703294,
	url          = {https://doi.org/10.1145/3613424.3623780},
	numpages     = 16
}

@inproceedings{makis2,
	title        = {Micro-MAMA: Multi-Agent Reinforcement Learning for Multicore Prefetching},
	author       = {Block, Charles and Gerogiannis, Gerasimos and Torrellas, Josep},
	year         = 2025,
	booktitle    = {Proceedings of the 58th IEEE/ACM International Symposium on Microarchitecture},
	location     = {},
	publisher    = {Association for Computing Machinery},
	address      = {New York, NY, USA},
	series       = {MICRO '25},
	pages        = {884–898},
	doi          = {10.1145/3725843.3756096},
	isbn         = 9798400715730,
	url          = {https://doi.org/10.1145/3725843.3756096},
	numpages     = 15
}

@article{casd_cal,
	title        = {Context-Aware Set Dueling for Dynamic Policy Arbitration},
	author       = {Patsidis, Diamantis and Vavouliotis, Georgios},
	year         = 2025,
	journal      = {IEEE Computer Architecture Letters},
	volume       = 24,
	number       = 2,
	pages        = {301--304},
	doi          = {10.1109/LCA.2025.3617159},
	url          = {https://doi.org/10.1109/LCA.2025.3617159}
}

@inproceedings{jamet2024practically,
	title        = {Practically Tackling Memory Bottlenecks of Graph-Processing Workloads},
	author       = {Jamet, Alexandre Valentin and Vavouliotis, Georgios and Jiménez, Daniel A. and Alvarez, Lluc and Casas, Marc},
	year         = 2024,
	booktitle    = {2024 IEEE International Parallel and Distributed Processing Symposium (IPDPS)},
	volume       = {},
	number       = {},
	pages        = {1034--1045},
	doi          = {10.1109/IPDPS57955.2024.00096},
	url          = {https://doi.org/10.1109/IPDPS57955.2024.00096}
}

@techreport{vavouliotis_free_tlb_prefetching,
	title        = {Pushing the Envelope on Free TLB Prefetching},
	author       = {Georgios Vavouliotis and Lluc Alvarez and Marc Casas},
	year         = 2021,
	url          = {https://upcommons.upc.edu/entities/publication/198eaf18-44e5-4ed3-8bfa-bd2f3e96e154},
	institution  = {Barcelona Supercomputing Center (BSC) and Universitat Polit{\`e}cnica de Catalunya (UPC)}
}

@phdthesis{vavouliotis_vm_prefetching,
	title        = {Advanced Hardware Prefetching in Virtual Memory Systems},
	author       = {Georgios Vavouliotis},
	year         = 2023,
	url          = {https://upcommons.upc.edu/entities/publication/f16d637b-ad15-4f69-b830-3471b7f2fb84},
	school       = {Universitat Polit{\`e}cnica de Catalunya (UPC)}
}
%%%%%%%%%%%%%%%%%%%%%%%%%%%%%%%%%%%%

\end{document}